\documentclass[final,5p,times,twocolumn,authoryear]{elsarticle}

\usepackage{aas_macros}
\usepackage{cite}
\usepackage{amsmath,amssymb,amsfonts}
\usepackage{algorithm}
\usepackage{algorithmic}

\usepackage{graphicx}
\usepackage{multirow}
\usepackage{xcolor}
\usepackage{lscape}
\usepackage{subfig}
\usepackage{float}
\usepackage{lipsum}
\usepackage{longtable}
\usepackage{xtab,booktabs}
\usepackage{verbatim}
\usepackage{hyperref}

\newcommand{\ie}{\emph{i.e.}}
\newcommand{\ovr}{One \emph{vs.} Rest}
\newcommand{\eg}{\emph{e.g.}}
\newcommand{\distclassipy}{\texttt{DistClassiPy}}
\newcommand{\dmc}{distance metric classifier}
\newcommand{\fone}{$F_1$}

\newtheorem{defn}{Definition}[section]
\newtheorem{cor}{Corollary}
\newtheorem{exmp}{Example}[section]

\usepackage{thmtools}
\usepackage{thm-restate}

\journal{Astronomy $\&$ Computing}

\begin{document}

\begin{frontmatter}

\title{Light curve classification with \distclassipy:\\ A new distance-based classifier}

\author[1,2]{Siddharth Chaini\corref{corrauth}}
\cortext[corrauth]{Corresponding author}
\ead{sidchaini@gmail.com}
\author[3,4]{Ashish Mahabal}
\author[5]{Ajit Kembhavi}
\author[1,6,7,8]{Federica B. Bianco}

\address[1]{Department of Physics and Astronomy, University of Delaware, Newark, DE  19716, USA}
\address[2]{Department of Physics, Indian Institute of Science Education and Research, Bhopal 462066, India}
\address[3]{Division of Physics, Mathematics and Astronomy, California Institute of Technology, Pasadena, CA 91125, USA}
\address[4]{Center for Data Driven Discovery, California Institute of Technology, Pasadena, CA 91125, USA}
\address[5]{Inter University Centre for Astronomy and Astrophysics (IUCAA), Pune 411007, India}
\address[6]{Joseph R. Biden, Jr. School of Public Policy and Administration, University of Delaware,   DE 19716,  USA}
\address[7]{University of Delaware, Data Science Institute, Newark, DE 19713, USA}
\address[8]{Vera C. Rubin Observatory, Tucson, AZ 85719, USA}

\begin{abstract}

The rise of synoptic sky surveys has ushered in an era of big data in time-domain astronomy, making data science and machine learning essential tools for studying celestial objects.
While tree-based models (\eg\ Random Forests) and deep learning models dominate the field, we explore the use of different distance metrics to aid in the classification of astrophysical objects. We developed \distclassipy, a new distance metric based classifier. The direct use of distance metrics is unexplored in time-domain astronomy, but distance-based methods can help make classification more interpretable and decrease computational costs.

In particular, we applied \distclassipy{} to classify light curves of variable stars, comparing the distances between objects of different classes.
Using 18 distance metrics on a catalog of 6,000 variable stars across 10 classes, we demonstrate classification and dimensionality reduction. Our classifier meets state-of-the-art performance but has lower computational requirements and improved interpretability. Additionally, \distclassipy{} can be tailored to specific objects by identifying the most effective distance metric for that classification. 

To facilitate broader applications within and beyond astronomy, we have made \distclassipy\ open-source and available at \url{https://pypi.org/project/distclassipy/}.

\end{abstract}

\begin{keyword}
Variable stars (1761) \sep Astronomy data analysis (1858) \sep Open source software (1866) \sep Astrostatistics (1882) \sep Classification (1907) \sep Light curve classification (1954)
\end{keyword}

\end{frontmatter}

\section{Introduction}

Over the last few decades, time-domain astronomy has experienced rapid growth.
This growth has been driven by the advent of large-scale
sky surveys like the Sloan Digital Sky
Survey~\citep[SDSS;][]{yorkSloanDigitalSky2000}, the Catalina Real-Time
Transient Survey~\citep[CRTS;][]{djorgovskiCatalinaRealTimeTransient2011} and
the Zwicky Transient Facility~\citep[ZTF;][]{bellmZwickyTransientFacility2019}
(see~\citealt{Djorgovski2013} for a comprehensive list of surveys), and accompanied
by advances in computing power and data storage. Observing billions of
astronomical objects over time allows us to detect changes in the night sky
that were once impossible to see. The Vera C. Rubin Observatory Legacy
Survey of Space and Time~\citep[LSST;][]{ivezicLSSTScienceDrivers2019}, will observe over 37 billion objects over its 10-year lifespan. However,
this opportunity for new discoveries comes with a
data-intensive challenge. Manual classification of all objects is
impossible,
so we need machine learning methods to automate the
classification and identification of objects of interest.

\begin{figure*}[t!]
    \centering
    \includegraphics[width=0.75 \paperwidth]{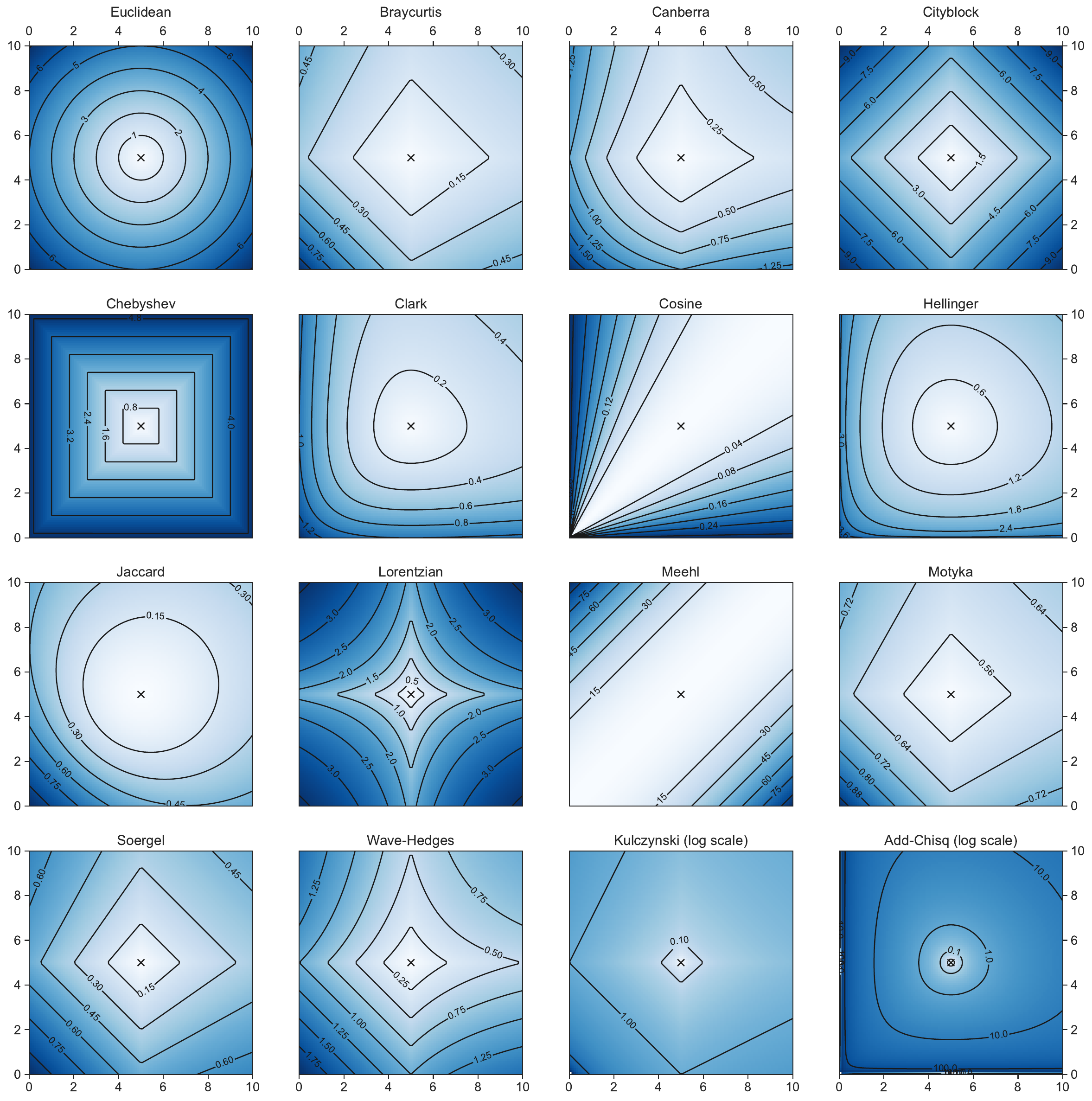}
    \caption{A visualization of 16 (of the 18) distance metrics used throughout this work. Each subplot shows the equidistant loci measuring the distance from the central point $(5,5)$. The color background denotes the distance values, with labeled contours. Contours  differ for each subplot, as the range of values the distance can take varies by metric. To aid readability, we use a log-scale for the last two metrics --- Kulczynski and Additive ChiSq due to high-power elements (in the metric definition) compressing the distance scale. The Correlation and Maryland Bridge metrics are not visualized here as they require vector inputs, and not 2-dimensional data points (see \ref{app:metrics}).}
    \label{fig:distances_viz}
\end{figure*}

Machine learning refers to a class of computer algorithms in which the computer learns patterns within the data, eliminating the need for explicit manual programming. Machine learning models can be deployed for a variety of cases --- for \eg, clustering (unsupervised learning) or for classification and regression tasks (supervised learning). In the context of large datasets and the need for automation, it has become essential in modern astronomy. 
A critical task in time-domain astronomy is classifying astronomical
objects based on how their brightness changes with time (called light curves),
which is well suited to be performed by machine learning algorithms (see for
example~\citealt{2002ASPC..259..160E,2002SPIE.4847..379T}
and~\citealt{2008AIPC.1082..287M}; for more recent work
see~\citealt{2020A&A...642A..58C}
and~\citealt{forsterAutomaticLearningRapid2020}). 
Because of the irregular sampling of astrophysical light curves, we generally extract features (such as statistical properties of the data, best-fit parameters to models, etc.) from the light curves, which are then fed to a classifier model. 
However, as the number of features increase, prediction
becomes difficult and computationally expensive, and performance typically
drops~\citep{bishopPatternRecognitionMachine2006}, making dimensionality
reduction a requirement. Furthermore, the use of more features may lead to data abstraction, and impair the
interpretability of a model - resulting in the model behaving like a black-box.

Dimensionality reduction involves mapping the higher-dimensional data into a lower-dimensional space while ensuring the low-dimensional representation retains the properties of the original data for prediction.

\begin{figure}
    \centering
    \includegraphics[width=\columnwidth]{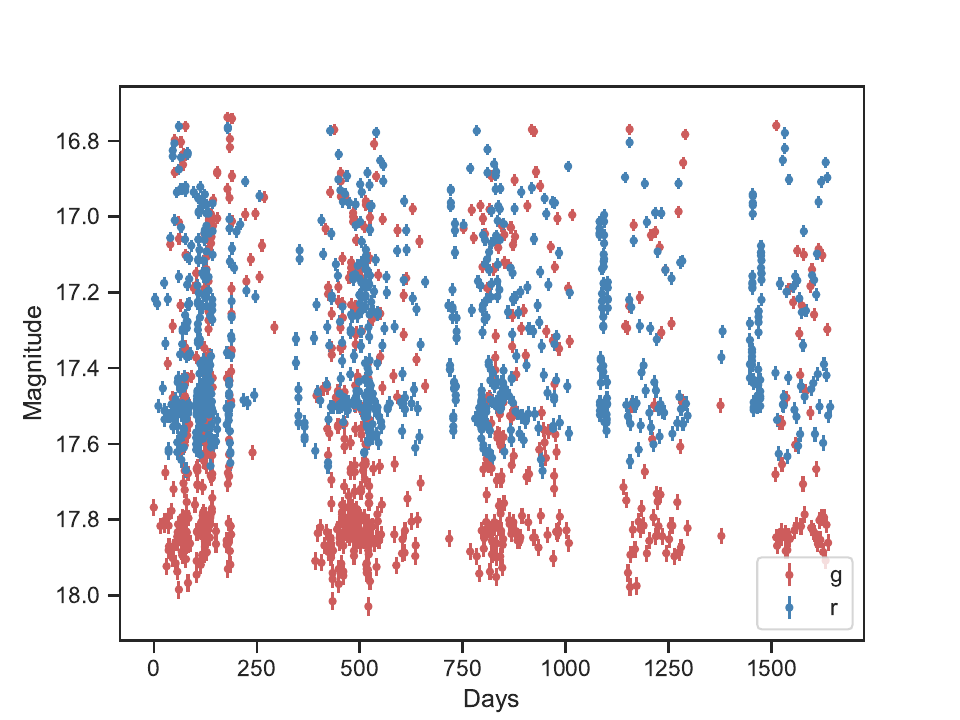}
    \caption{An example light curve of a RR Lyrae Type ab (RA~=~2.93, Dec~=~44.62, period~=~0.55 days), taken from the ZTF DR15. Each light curve consists of a series of magnitude (brightness) values as a function of time, which in our case is for the two filters --- $g$ and $r$. The sampling of ZTF is sparse, as common for ground-based surveys, with visible gaps due to seasonality, and, although this is a periodic variable, the periodicity is not obvious because of the sampling.}
    \label{fig:lc}
\end{figure}

In a physical space, a distance is a scalar quantity that tells us how far away two objects are from each other. In a feature space, the ``distance'' is inverse to the similarity of the objects. A distance metric\footnote{Also referred to as a distance measure in literature.} refers to the mathematical function or algorithm used to measure the distance between two points in a space. Using the appropriate metric and features, we expect that the distance between objects of the same class should be smaller than between objects from different classes. These distances can then be used to separate and classify light curves.

In a physical space, a distance is a scalar quantity that tells us how far away two objects are from each other. In a feature space, the ``distance'' is inverse to the similarity of the objects. A distance metric\footnote{Also referred to as a distance measure in literature.} is the mathematical function or algorithm used to measure the distance between two points in a space. Using the appropriate metric and features, we expect that the distance between objects of the same class should be smaller than between objects from different classes. These distances can then be used to separate and classify light curves.

In this paper, we compare the effectiveness of different
distance metrics applied for dimensionality reduction and classification of
light curves. Using a wealth of distances defined in statistics
and
mathematics~\citep{chaComprehensiveSurveyDistance2007,dezaEncyclopediaDistances2013,tschopp2017quantifying},
we compiled a list of 18 distance metrics designed for data analysis,
that have been defined in the appendix, 
and visualized in \autoref{fig:distances_viz}. We use these metrics to compare classification performance for three classification problems, described in \autoref{sec:classification}. We also identify the most important features for each distance metric and classification task, and analyze the impact of limiting the feature space to these top features, thus reducing dimensionality and computational cost. The use of different distance metrics for dimensionality reduction and subsequent classification is an approach that, to our knowledge, has not been explored in time-domain astronomy before. 

We show that our classifier, \distclassipy, when used natively out of the box,
offers similar performance to other models commonly used to classify light curves, but with lower computational costs. In addition, our model can be
tailored according to scientific interests by knowledgeably selecting the
distance and features to use based on a scientific goal and data
characteristics, enhancing both the computational and performance properties of the
model. This paper is accompanied by a Python package,
\distclassipy,\footnote{\url{https://pypi.org/project/distclassipy/}} which
is our \dmc\ built on top of \texttt{scikit-learn}~\citep{scikit-learn}, in
addition to all code required to reproduce our results in this
paper.\footnote{\url{https://github.com/sidchaini/LightCurveDistanceClassification}}

We introduce some of the scientific and technical background in \autoref{sec:ml_intro}, describe our data in \autoref{sec:data}, discuss our approach to dimensionality reduction in \autoref{sec:feature_selection}, then our approach to perform and evaluate the performance on our classification tasks (\autoref{sec:classification}). We describe our results in \autoref{sec:results} and conclude in \autoref{sec:conclusions}.

\begin{figure*}[ht!]
    \centering
    \subfloat[Before outlier removal.]{
        \label{subfig:Cep_before}
        \includegraphics[width=0.49\textwidth]{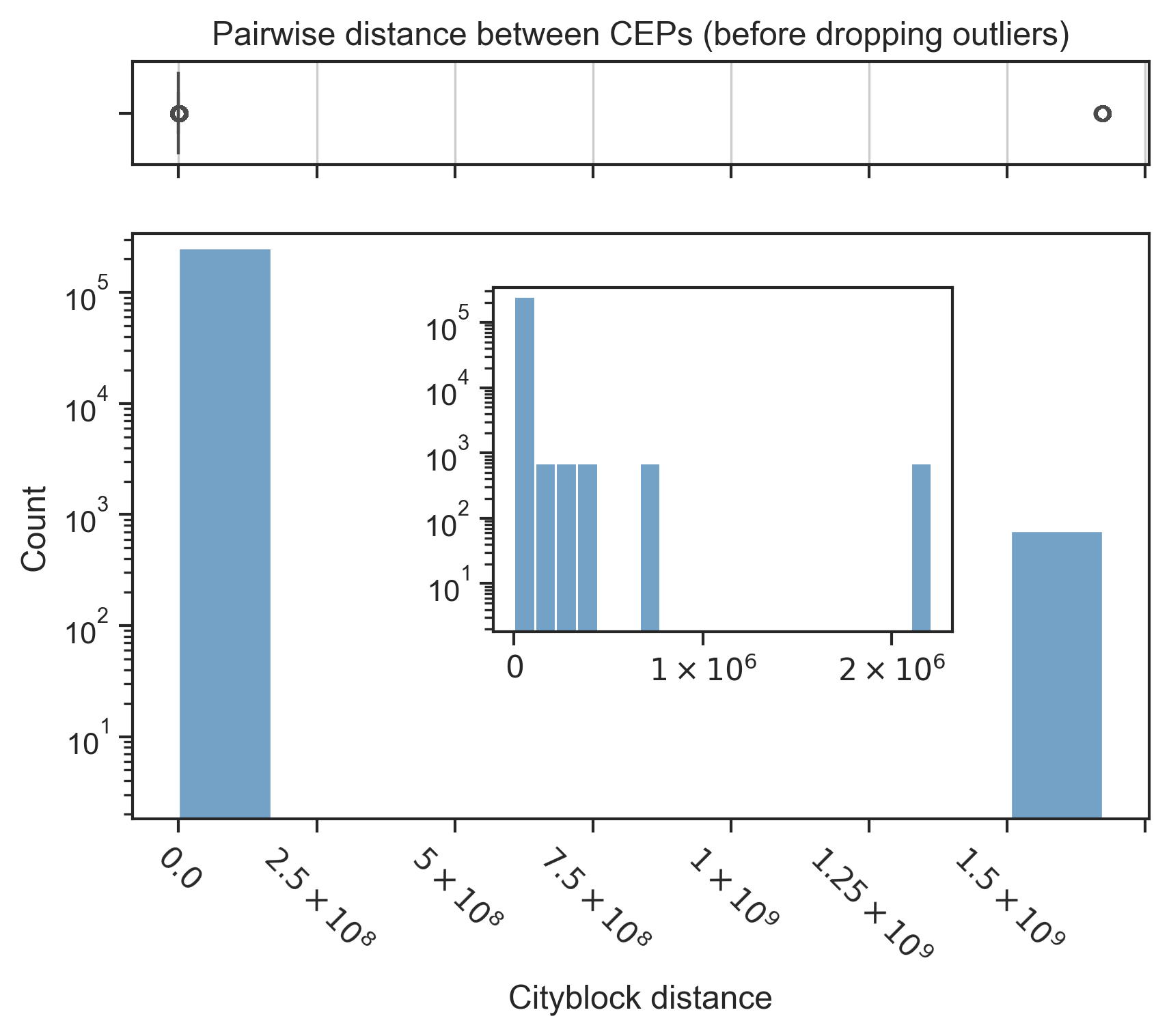}
    }
    \subfloat[After outlier removal]{
        \label{subfig:Cep_after}
        \includegraphics[width=0.49\textwidth]{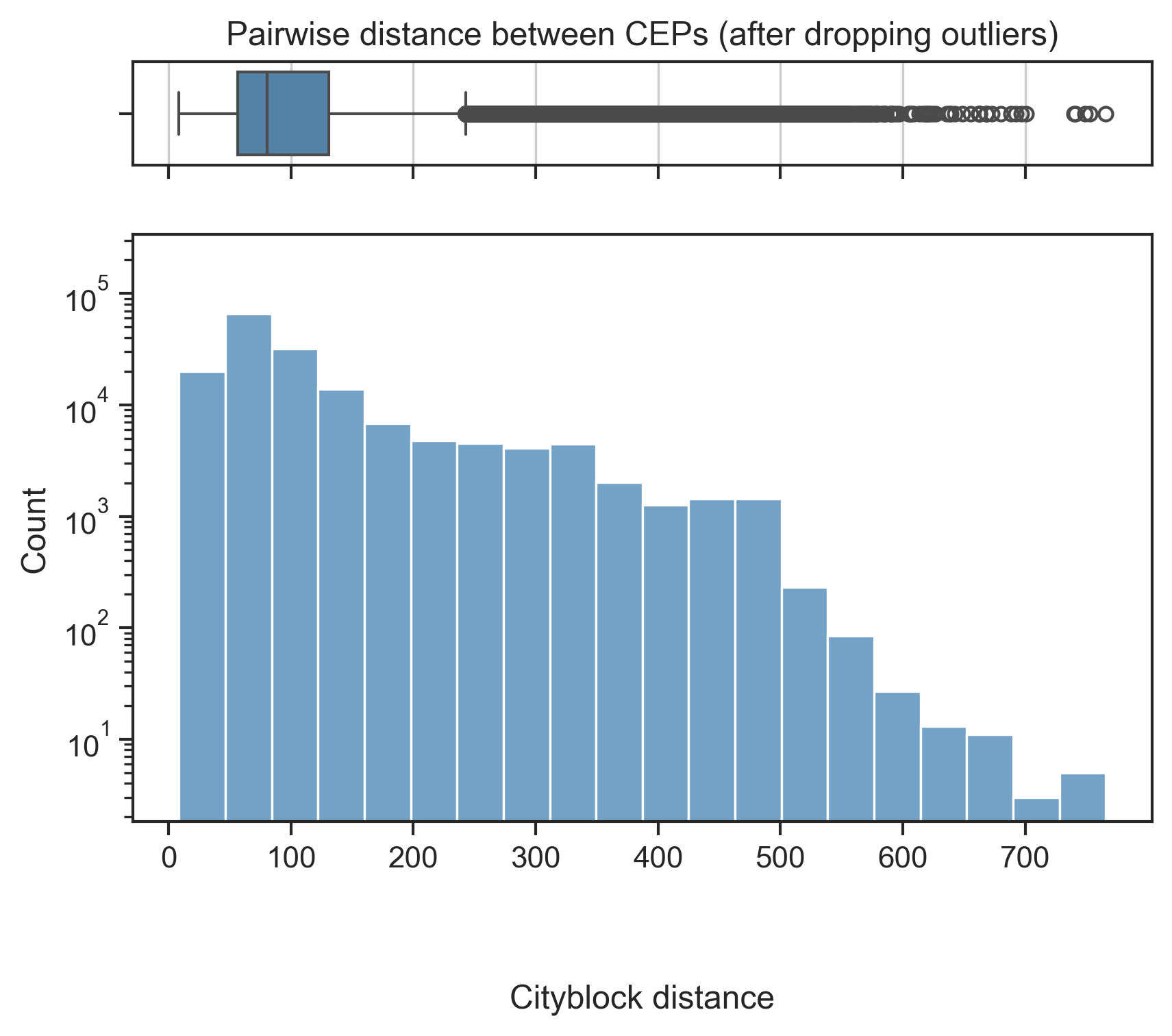}
    }
    \caption{The distribution of pairwise Cityblock distances ($d_{CB}$) between all 558 Cepheid variable stars in our final dataset (class CEP): 155,403 unique pairs of CEP contribute one point each to this distribution. The left and right panels show the distribution of distances before and after outlier removal respectively.  The top panels show a box and whiskers plot of the same, where the median is marked by a vertical line, the interquartile range by the box, the 10th and 90th percentile by the whiskers. All points beyond these percentile values (our definition of outlier, see \autoref{subsec:data_cleaning}) are plotted individually. In the left plots (before outlier removal) the original distribution spreads out to $d_{CB}>10^9$ but its extremely sparse past $d_{CB}\geq10^5$ (see insert which zooms into the $d_{CB}<10^6$ region, and notice how in the top plot the box and whiskers are indistinguishable). However, outlier removal leads to a much more compact distribution of pair-wise distances (right) where the distances are contained between $0 <d_{CB} <1,000$. Note that, as the distribution changes after the first cut, the functional definition of outliers does not, thus the tail of the distribution is plotted with individual points in the right-side figure.}
    \label{fig:outlier_removal}
\end{figure*}

\section{Distances in Machine Learning}\label{sec:ml_intro}
The concept of distance is intuitive. It tells us about the degree of proximity between two objects --- the shorter the distance, the closer the objects. However, the distance depends on the path followed to join two objects. Because of this, we can have different types of distances, each calculated differently. Let us first define this mathematically:
\begin{restatable}{defn}{defdistance}
\label{def:distance}
The distance $d$ between two points, in a set $X$, is a function $ d: X \times X \rightarrow [0, \infty)$ that gives a distance between each pair of points in that set such that, for all $x,y,z \in X$, the following properties hold:
\begin{enumerate}
\item $d(x,y) = 0 \iff x = y$ (identity of indiscernibles)
\item $d(x,y) = d(y,x)$ (symmetry)
\item $d(x,y) \leq d(x,z) + d(z,y)$ (triangle inequality)
\end{enumerate}\end{restatable}

An example of a distance is the Euclidean distance, commonly used for measuring the distance between physical objects (the ``as the crow flies'' distance). 
A detailed discussion of distance metrics, along with the definition of each of the 18 metrics we used, are included in the appendix, \ref{app:metrics} and \ref{app:list_of_metrics}, respectively. Of these 18 distances, 16 are visualized in a 2-dimensional space in \autoref{fig:distances_viz}.

Distance metrics see their use in a variety of supervised and unsupervised
machine learning algorithms (\eg, $k$-Nearest Neighbors
---~\citealt{abualfeilatEffectsDistanceMeasure2019,nayakStudyDistanceMetrics2022} ---, Kernel Density Estimation
---~\citealt{heKernelDensityMetric2013} ---, hierarchical clustering
---~\citealt{murtagh2012algorithms}).

In many machine learning tasks, we do not feed the data directly to a machine learning model. Instead, we extract or engineer \emph{features} from the data and transform them into a new space, to which we will refer hereafter as the ``feature space''. The dimensionality of the feature space is determined by the number of features, denoted as $n$. The feature space itself exists as a real coordinate space, represented as $\mathbb{R}^n$. 
In this $n$-dimensional feature space, we can now define a variety of distance metrics: any function complying with \autoref{def:distance} is a valid distance metric. When equipped with a distance metric, the feature space takes on the characteristics of a metric space (as defined in \ref{def:metric_space}). Since the distance value is always a positive real number, we can easily compare this value for different points even when dealing with high-dimensional data.

\section{Data} \label{sec:data}
\subsection{Catalog and Raw Light Curves} \label{subsec:catalog}
Our dataset consists of light curves of variable stars from the Zwicky
Transient Facility (ZTF) Data Release 15
(DR15).\footnote{\url{https://irsa.ipac.caltech.edu/data/ZTF/docs/releases/dr15}}
ZTF is a robotic synoptic facility located at the Palomar Observatory. It scans
the sky, monitoring objects of magnitude,\footnote{The magnitude scale is a
logarithmic brightness scale defined for astrophysical objects where the
brightness of an object with flux $F$ is measured in magnitudes
$m$ as $m = -2.5 \log_{10}(F/F_0)$ and where $F_0$ is an instrumental
normalization factor.} $r <=20.6$~\citep{bellmZwickyTransientFacility2019}.
ZTF observes in three wavelength bandpasses, $g$, $r$ and
$i$
in the optical wavelength regime. However, the $i$ band has a lower cadence as well as shallower observations, and so we restrict our data to the $g$ and $r$ bands. In our dataset, there are on average 382 points in the $g$ band and 674 points in the $r$ band for each light curve, which are collected over an average of 4.42~years (minimum 0.61~years, maximum 4.64~years, median 4.47~years). 

The original catalog \citep{chenZwickyTransientFacility2020a} consists of 781,602 identified variable stars in 10 classes (see \autoref{tab:classes}) that represent stellar objects of different nature and that display different observational properties in the time domain. Examples of light curves in our dataset are shown in \autoref{fig:lc} 
The original catalog~\citep{chenZwickyTransientFacility2020a} consists of
781,602 identified variable stars in 10 classes (see \autoref{tab:classes}) that
represent stellar objects of different nature and that display different
observational properties in the time domain. Examples of light curves in our
dataset are shown in \autoref{fig:lc}. 
We will briefly return to the topic of variable stars classification in
\autoref{sec:classification}; for a comprehensive review of variable stars
classification and characterization, see~\citet{eyer2008variable}. The number
of objects in each class in~\citet{chenZwickyTransientFacility2020a} is heavily
imbalanced, with the catalog containing 369,707 Eclipsing W Ursae Majoris (EW)
variables, but only 1,262 Cepheid (CEP) variables. 
Models trained on an imbalanced dataset have a hidden bias due to the relative frequencies of occurrence which teaches the model that predictions on minority classes carry a significant risk 
thus impacting the performance on these
classes~\citep{krawczykLearningImbalancedData2016}. Thus, we select a random
set of 1,000 objects of each class
from~\citet{chenZwickyTransientFacility2020a}'s dataset to develop our model. 
Our raw dataset consists of 10,000 light curves (1,000 for each of the 10 classes listed in \autoref{tab:classes}). After cleaning the data and dropping outliers (as described in \autoref{subsec:data_cleaning}), we are left with 558 objects from each of the ten classes.

In addition to the above data, we also used an entirely ``hidden'' set of $\sim$500 new objects for each class as a final test set. The results of these final performance tests are discussed in \autoref{sec:conclusions}.

\subsection{Feature Extraction} \label{subsec:feature_extraction}
Since the ZTF is a ground-based telescope, the light curves obtained are unevenly sampled, sparse, noisy, and heteroskedastic. 
Objects from the same class have different noise levels and different magnitudes based on their distance from the Earth. This extrinsic diversity within a class makes the direct comparison of light curves difficult. 
Our scheme for feature extraction is domain-driven: we select features that measure expected behaviors of variable stars (\eg, periodicity). 

To extract features from the light curves, we use the \texttt{lc\_classifier}
module~\citep{jainagaAlercebrokerLcClassifier2021} in Python. While the context
of the astrophysical target may be informative (\eg, its coordinates to
indicate if it is likely to be a Galactic or extra-galactic object), all of
\texttt{lc\_classifier}'s features are based only on the light curve data. Most
features (53) are calculated separately for the $g$ and $r$
passband, while some features (8) are calculated jointly from the $g$
and $r$ observations. 
This gives us a total of 114 features for every light curve in our dataset. 

A detailed description of all the features is provided in~\citet[][see Table~2]{sanchez-saezAlertClassificationALeRCE2021}. Most features are based on
common light curve statistics (\eg\ amplitude --- based on the difference of
highest and lowest magnitudes), while some are based on parameters obtained
after fitting a model to the light curve (\eg\ multiband period --- the period is
obtained by fitting a periodogram).

\subsection{Data Cleaning} \label{subsec:data_cleaning}
We remove objects from our dataset for which feature extraction failed for one
or more features. For most classes, this removes 3\%--8\% of the objects.
However, we find that the classes Mira, CEP, and SR have a higher fit failure
rate for some features like $\eta$ (Ratio of mean of square of successive
mag differences to light curve variance;
refer~\citet[][Table~3]{kimEPOCHProjectPeriodic2014}) and the Mexican Hat Power
Spectra~\citep{arevaloMexicanHatHoles2012}, thus leaving us with only 640
Miras, 713 CEPs, and 741 SRs. 
To keep the dataset balanced, we once again randomly drop objects from each class such that each class has exactly 600 objects. However, we do \emph{not} rebalance the hidden set of data --- this is to obtain a more representative assessment of our model's performance.

\citet{chenZwickyTransientFacility2020a} state that their catalog does not
constitute a robust classification, 
thus we expect our labels to be noisy and that
there might be outliers in the dataset.

\begin{figure*}[t!]
    \centering
    \includegraphics[width=0.6\textwidth]{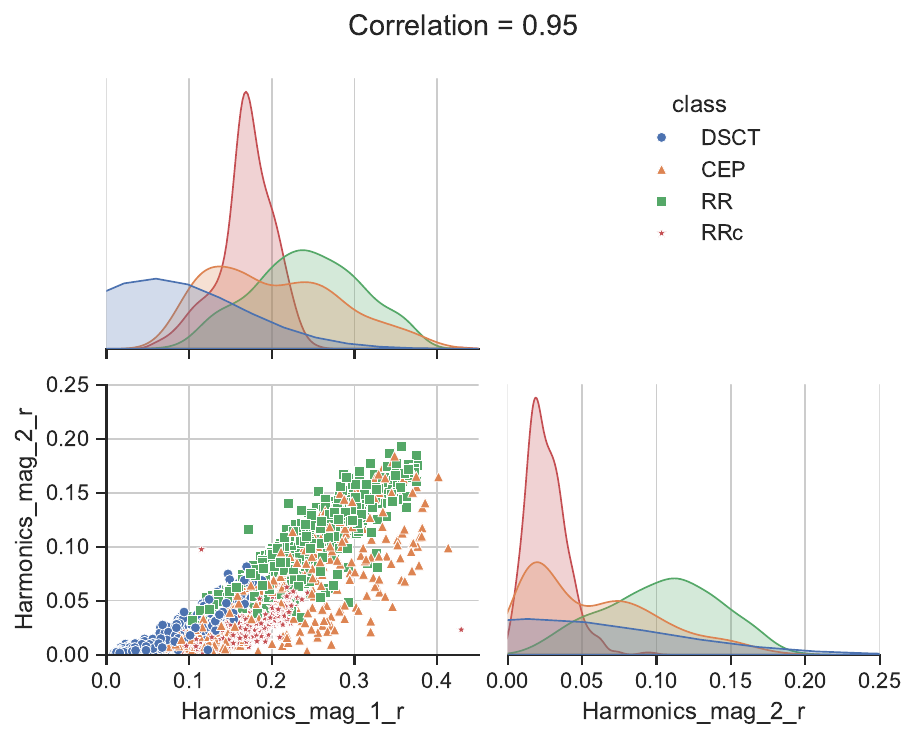}
    \caption{Correlation between the first two components of a harmonic series fit to each light curve in the $r$-band. The bottom left panel shows the linear relationship between the two components, while the remaining two panels illustrate the distribution of the component per class. We find that these two components have a Pearson's linear correlation coefficient $r=0.95$, and thus choose only \texttt{Harmonics\_mag\_1} as part of our feature selection step.}
    \label{fig:harmonics_corr}
\end{figure*}

To assess the presence of outliers 
we calculate the pair-wise distances with the Cityblock metric in the 114-dimensional feature space among all classes. 
We remove the top 10\% and bottom 10\% of the distribution to eliminate outliers and misclassifications, as well as potential duplicates in the ZTF photometry.  The process has been illustrated for the class CEP in \autoref{subfig:Cep_before}. We see a prominent peak centered near 0 and a long tail that extends to $2.13 \times 10^6$ in the original distribution. After the cuts, the distribution of distances remains right-skewed, with a long tail. 
The distribution of pairwise Cityblock distances for CEPs after outlier removal is also illustrated in \autoref{subfig:Cep_after}. This step is largely robust to the choice of distance metric used (\eg, the Euclidean and Cityblock metrics lead to datasets with an overlap of $>$98\%).

Finally, we have a `clean' dataset of 558 objects of each class mentioned in \autoref{tab:classes}.

\begin{table}
    \centering
    \resizebox{\columnwidth}{!}{%
        \begin{tabular}{p{2cm} p{3cm} p{2cm} p{1.5cm}}
            \hline
    		\textbf{Abbreviation} & \textbf{Class Name}  & \textbf{Original class size} & \textbf{Final class size}     \\ \hline
    		BYDra                 & BY Draconis               &84,697 &.     \\
      CEP                   & Cepheid         & 1610 & .          \\
    		DSCT                  & Delta Scuti              & 16,709 & . \\
    		EA                    & Eclipsing Algol          & 49,943	 & . \\
    		EW                    & Eclipsing W Ursae Majoris  & 369,707 & 558\\
    		Mira                  & Mira                   & 11,879 & .   \\
    		RR                    & RR Lyrae (Type ab)       & 32,518 & .\\
    		RRc                   & RR Lyrae Type c         & 13,875  & .\\
    		RSCVN                 & RS Canum Venaticorum     & 81,393 & . \\
    		SR                    & Semiregular              & 119,261 & . \\ \hline
        \end{tabular}
        }
\caption{Correlation between the first two components of a harmonic series fit to each light curve in the $r$-band. The bottom left panel shows the linear relationship between the two components, while the remaining two panels illustrate the distribution of the component per class. We find that these two components have a Pearson's linear correlation coefficient $r=0.95$, and thus choose only \texttt{Harmonics\_mag\_1} as part of our feature selection step.}
\label{tab:classes}
\end{table}

\section{Feature Selection and Dimensionality Reduction} \label{sec:feature_selection}

We are up to this point working in a 114-dimensional feature space. Working in
high-dimensional spaces may hinder the performance of machine learning models due to the
``curse of dimensionality''~\citep{bishopPatternRecognitionMachine2006}, which
leads to a decrease in performance and an increase in computational complexity.
To address this, we further reduce the dimensionality of the 114 feature space
using a variety of approaches. We refer to this step as feature selection,
because we are selecting a subset of features from those created in
\autoref{subsec:feature_extraction}.

\subsection{Drop all g-band features}
Our data consists of light curves measured in two wavelength passbands, $g$ and $r$. We used \texttt{lc\_classifier} to extract the aforementioned features from each band, and from the ``multiband'' light curve generated as the difference between the two, which represents the color of the transient.\footnote{Since magnitude is a log scale, the difference of the magnitudes is the ratio of brightness in the two bandpasses.}
Because the multiband features contain information from both $g$ and $r$ bands, we dropped all $g$-band features as a dimensionality reduction step. By removing only $g$-band features, we are not removing any information, since it is indirectly contained in the multiband features derived from both the $r$ and $g$ bands. We chose to remove $g$-band features instead of $r$-band features because ZTF is more sensitive in the $r$-band, and also has more observations in the $r$-band. This step reduces the dimensionality of the feature space from 114 to 60.

\subsection{Dropping flags and number of points}
We dropped features that are not physically motivated or are not well suited to measure distances. These include the number of points in a light curve,\footnote{Note that we  already remove light curves with few points to avoid introducing bias in  feature extraction.} and flags regarding the success of models fits (because of the low dynamic nature of their binary values). This removes 16 features.

\subsection{Dropping Highly Correlated Features}

Features having high correlation (which we measure with Pearson's linear correlation coefficient, $r$) do not add much new information to the classification. So, for every set of highly correlated features (\ie, $r>0.9$)\footnote{The threshold $0.9$ was empirically found to be suitable to balance the trade-off between reducing redundancy and retaining informative features.},  we keep only one from that set.
Features having high correlation (which we measure with Pearson's linear correlation coefficient, $r$) do not add much new information to the classification. So, for every set of highly correlated features (\ie, $r>0.9$),\footnote{The threshold 0.9 was empirically found to be suitable to balance the trade-off between reducing redundancy and retaining informative features.} we keep only one from that set.

For example, \autoref{fig:harmonics_corr} shows the correlation between two features for the case of multi-class classification (see \autoref{sec:classification}) --- the amplitude of the first two components of a harmonic series fit to each light curve in the $r$-band, for a subset of our 10 classes. Because this correlation is high ($r>0.9$), we only choose one of these features (\texttt{Harmonics\_mag\_1}).

\begin{figure}
    \centering
    \subfloat[Before feature selection.]{
        \label{subfig:corr_matrix_before}
        \includegraphics[width=0.9\linewidth]{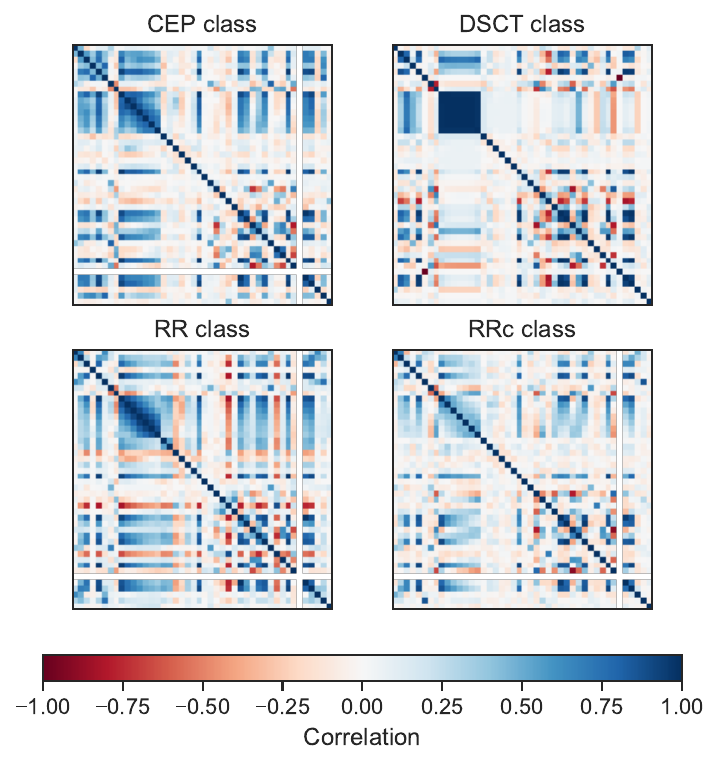}
    }
    \\
    \subfloat[After feature selection.]{
        \label{subfig:corr_matrix_after}
        \includegraphics[width=0.9\linewidth]{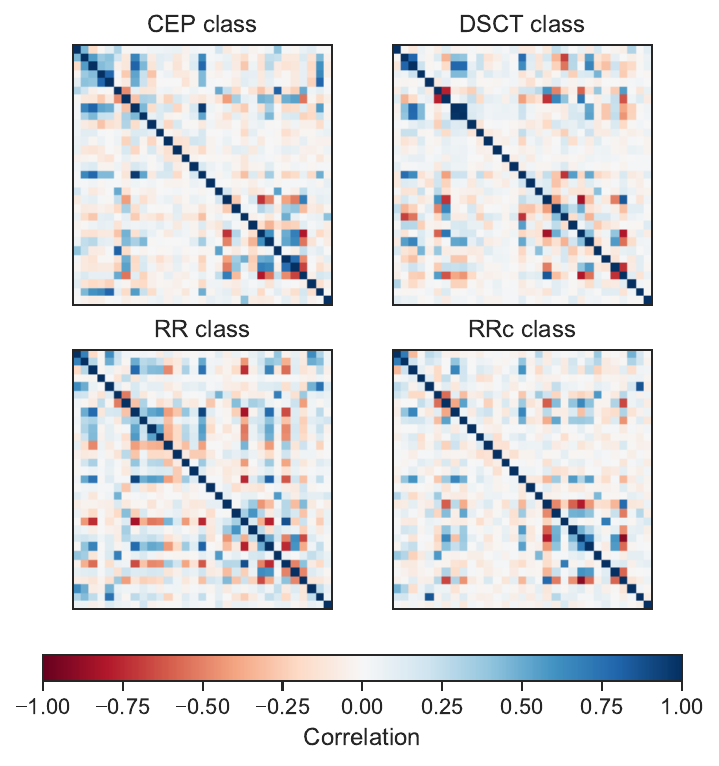}
    }
    \caption{Correlation matrices of our features space before and after dimensionality reduction for four exemplary labels: CEP, DSCT, RR, RRc (see \autoref{tab:classes} for a definition of each class), which we use for the multi-class classification problem. Each cell in the plot represents the Pearson's linear correlation coefficient, $r$, between the corresponding feature on the $x$ and $y$ axis. By definition, the diagonal has a correlation $r=1$. No correlation ($r=0$) is mapped to the color white, positive correlation ($r>0$) to shades of blue, while negative correlation ($r<0$) to shades of red. Our original set of 114 features per object (a) is computed using \texttt{lc\_classifier}. These consisted of features calculated in the $r$, $g$, and $g-r$ (multiband) light curves. In (a), the correlation between features is evident for each class in the block structure of the correlation matrix. We dropped $g$-based features, features that are not physically motivated, and finally, features that have a very high correlation. This leads to the correlation matrices shown in (b). Our final feature space has 31 features.}
    \label{fig:corr_matrix}
\end{figure}

After dropping the features as described above, we are left with 31 of the 114 feature dimensional space. A correlation matrix of the original 114 features and of the final set of 31 features is shown in 
\autoref{fig:corr_matrix}.

\section{Classification} \label{sec:classification}

\subsection{Classification Problems} \label{subsec:classification_problems}

Variable stars, and most stars which vary to some degree, are classified based
on their observational features into classes. Their variability is generally
connected to physical properties such as mass, metallicity (or chemical
composition), age, and environment, with the goal of understanding their
observational properties, including their variability, as an expression of
physical processes within the star (or star system in the case, for example, of
eclipsing binaries). A wealth of classes have been identified in astrophysical
studies~\citep{eyer2008variable}. 
To demonstrate the potential of our classifier in the classification of variable stars' light curves, 
we examine three typical, progressively more complex classification scenarios:

\begin{enumerate}
    \item \ovr\ Classification: EA, rest\footnote{Note: Rest includes all other classes combined.}
    \item Binary Classification: RSCVN, BYDra
    \item Multi-class Classification: CEP, RR, RRc, DSCT
\end{enumerate}

\begin{figure*}
    \centering
    \includegraphics[width=\columnwidth]{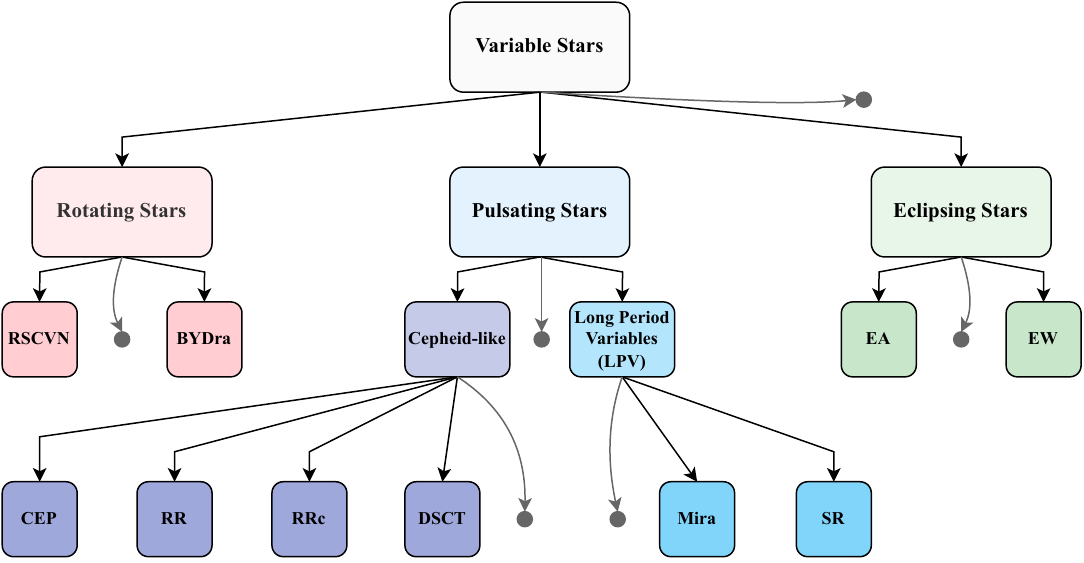}
    \caption{The canonical hierarchy tree for the ten classes included in our work.
    This is a physical classification, based on~\citet{eyer2008variable}.
    Note: the gray dot represents the objects which belong to none of the
    other classes, which are not represented in our
    dataset.}
    \label{fig:hierarchy}
\end{figure*}

For each classification task, we also run a Random Forest
Classifier~\citep[RFC;][]{breimanRandomForests2001},  a well-known machine
learning model which uses ensembles of decision trees models to classify a test
object. We choose RFC as our benchmark comparison because most astrophysical
services use RFC as the (or one of the) classifiers of choice
(\eg~\citealt{hinnersMachineLearningTechniques2018,sanchez-saezAlertClassificationALeRCE2021,cheungNewClassificationModel2021}). 
\autoref{fig:hierarchy} shows the canonical hierarchical classification scheme for
the ten classes included in our work, based
on~\citet[][Figure~1]{eyer2008variable}. Note that this is a physical, rather
than a phenomenological classification, so that, while in general they display
marked similarity, proximity in this classification scheme does not necessarily
imply proximity in the photometric feature space.

Our analysis will focus on the more comprehensive multi-class classification, where we look at four classes of Cepheid-like pulsating variables: Cepheid (CEP),  RR Lyrae Type ab (RR), RR Lyrae Type c (RRc), and Delta Scuti (DSCT). We discuss in detail these results in \autoref{sec:results}. But we also report the performance of our method on a specific binary case and \ovr\ case for comparison, keeping in mind that the results obtained when choosing one or two classes may differ significantly for any other class or class combination. In the \ovr\ classification, we isolate Eclipsing Algols (EA) from all other classes. In the binary classification, we separate two classes of rotating variables (RS Canum Venaticorum, RSCVN, and BY Draconis, BYDra).  This is a particularly challenging classification because while physically different, both classes exhibit photometrically similar behaviors, with  variability at the 
$\sim0.1 - 0.2$~mag level, and with varying phase, magnitude slope, amplitude, and
period~\citep{bopp_1980}.

\subsection{\distclassipy\ Classification Algorithm} \label{subsec:classification_algorithm}
We use a custom algorithm, which we have named \distclassipy, to classify the
light curves. Our method draws inspiration from the $k$-Nearest
Neighbors
($k-$NN;~\citep{coverNearestNeighborPattern1967,fixDiscriminatoryAnalysisNonparametric1989})
algorithm. In addition to the classification that $k-$NN provides,
\distclassipy\ also offers a quantification of the uncertainty in the
prediction.
We have implemented this in Python using the \texttt{scikit-learn}
API~\citep{sklearn_api}.

The detailed training algorithm is outlined in Algorithm \ref{alg:train}.

\subsubsection{Training}

\begin{table}[t!]
    \centering
    \renewcommand{\arraystretch}{1.15}
        \begin{tabular}{lllll}
            \hline
            Class & $r_\text{period}$ & $r_\text{amplitude}$ & $r_\text{R21}$ & $r_\text{skew}$ \\ \hline
            CEP            & 3.86                 & 0.25              & 0.27        & -0.16        \\
            RR             & 0.56                 & 0.35              & 0.45        & -0.52        \\
            RRc            & 0.33                 & 0.20              & 0.15        & 0.01         \\
            DSCT           & 0.09                 & 0.08              & 0.17        & -0.14        \\ \hline
        \end{tabular}

\caption{An example of the median set for 4 $r$-band features and 4 classes from the multi-class classification case. This is just a subsection of the complete data, the full table is available online at \url{https://github.com/sidchaini/LightCurveDistanceClassification}.}

\label{tab:median_set}
\end{table}

We compute the median and standard deviation for each feature per class, where we use the median set as a representative for each class.\footnote{The median and standard deviation can also be replaced by other statistical measures of central tendency and dispersion.} Using the median here also makes our calculations more resilient to outliers. An illustration of the median values for four representative features can be found in \autoref{tab:median_set}. The distance metric is not selected in the training step.

\begin{algorithm}
\caption{Training Step}
\label{alg:train}
\begin{algorithmic}[1]
\FOR{each class $C$ in the training set}
    \FOR{each feature $F$ in class $C$}
        \STATE Calculate the median $M_F^C$ of feature $F$ in class $C$.
        \STATE Calculate the standard deviation $\sigma_F^C$ of feature $F$ in class $C$.
    \ENDFOR
    \STATE Save the median set $\{M_F^C\}$ and standard deviation set $\{\sigma_F^C\}$ for class $C$.
\ENDFOR
\end{algorithmic}
\end{algorithm}

\begin{figure*}[t!]
    \centering
    \subfloat[Clark metric]{
        \label{subfig:clark_sfs}
        \includegraphics[width=0.49\textwidth]{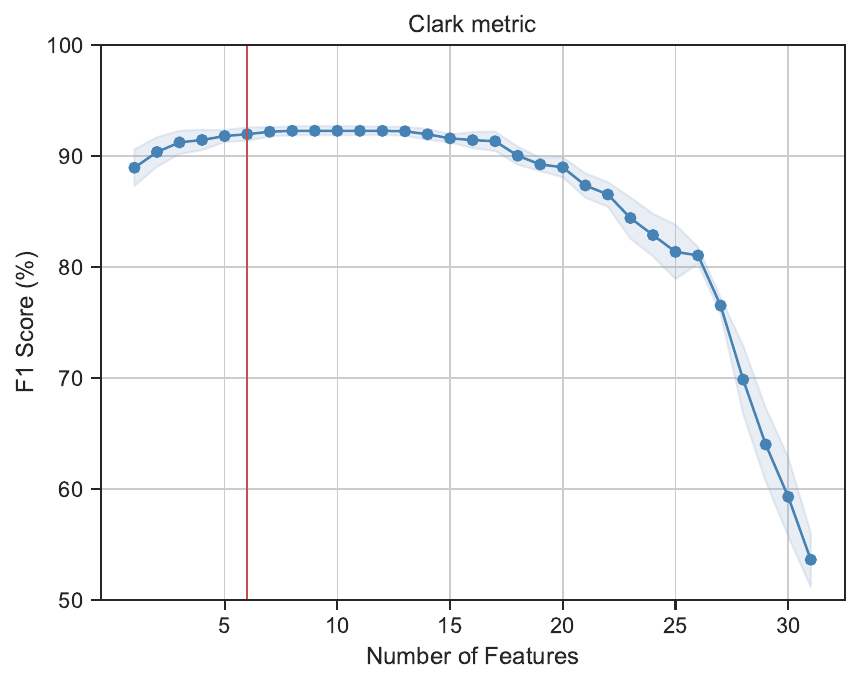}
    }
    \subfloat[Canberra metric]{
        \label{subfig:canberra_sfs}
        \includegraphics[width=0.49\textwidth]{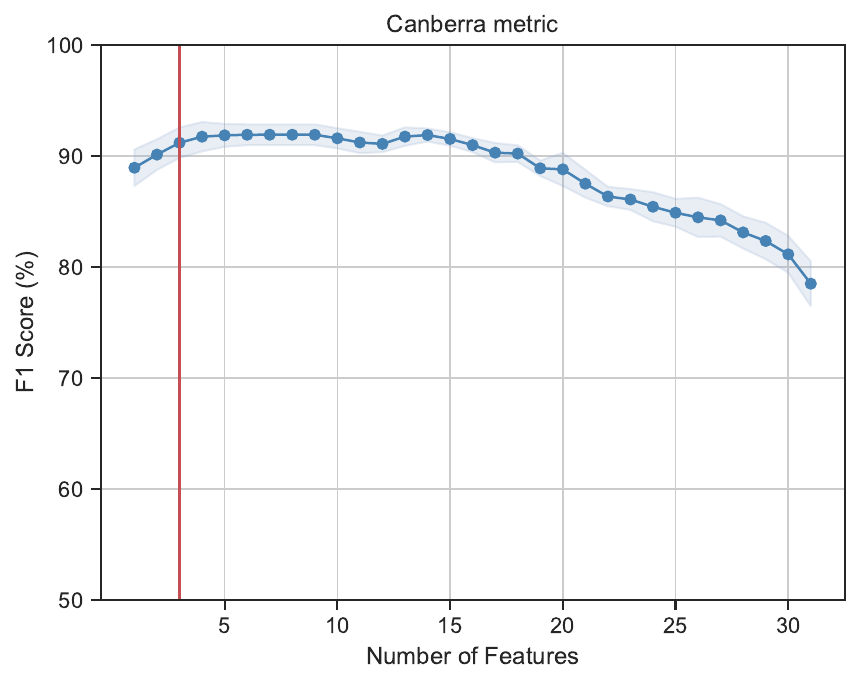}
    }
    \caption{\distclassipy\ permutation feature importance for the multi-class classification task for all metrics. We only show the importance for features that were selected for the classification with at least three metrics (see \autoref{fig:dcpy_topfeats}), and all other features are shown in gray. The \texttt{Period\_band\_r} appears as important for nearly all metrics (but not for the Maryland Bridge metric, which, as discussed in \autoref{fig:distances_viz} and \ref{app:metrics}, is a vector-based metric). A negative importance (\eg, \texttt{Psi\_Eta\_r} with Chebyshev) denotes the fact that, although using the SFS it was selected as one of the top four best features, when using all 31 features, it led to a decrease in performance. The Cosine and Correlation metrics stand out, with classification importance nearly entirely placed on the \texttt{GP\_DRW\_tau\_r} feature: the relaxation time for a damped random walk model of the light curve, originally designed in~\citet{2017MNRAS.470.4112G} for quasar classification.}
    \label{fig:sfs_plots}
\end{figure*}

\begin{figure*}[t!]
    \centering
    \includegraphics[width=0.85\textwidth]{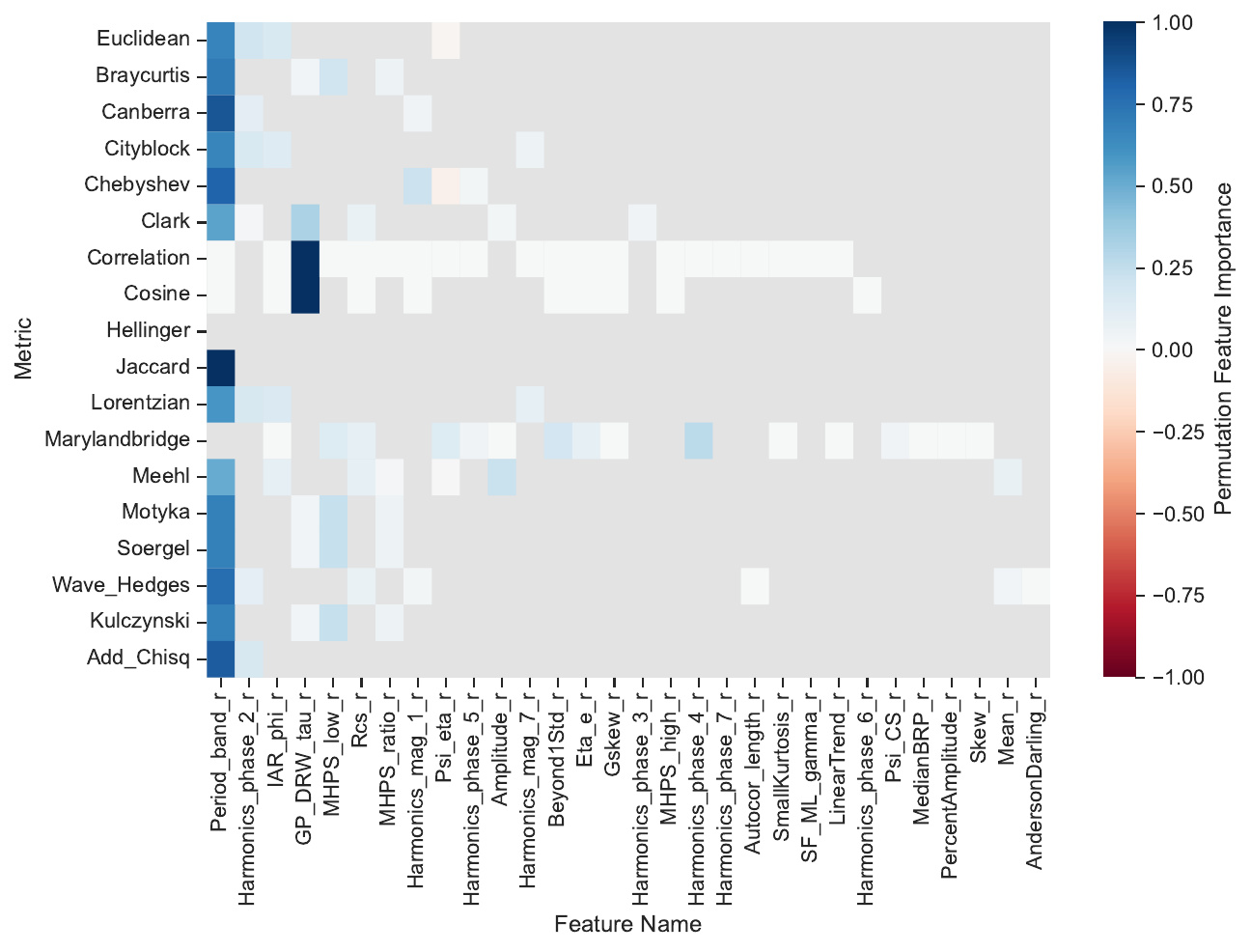}
    \caption{\distclassipy\ permutation feature importance for the multi-class classification task for all metrics. We only show the importance for features that were selected for the classification with at least three metrics (see \autoref{fig:dcpy_topfeats}), and all other features are shown in gray. The \texttt{Period\_band\_r} appears as important for nearly all metrics (but not for the Maryland Bridge metric, which, as discussed in \autoref{fig:distances_viz} and \ref{app:metrics}, is a vector-based metric). A negative importance (\eg, \texttt{Psi\_Eta\_r} with Chebyshev) denotes the fact that, although using the SFS it was selected as one of the top four best features, when using all 31 features, it led to a decrease in performance. The Cosine and Correlation metrics stand out, with classification importance nearly entirely placed on the \texttt{GP\_DRW\_tau\_r} feature: the relaxation time for a damped random walk model of the light curve, originally designed in~\citet{2017MNRAS.470.4112G} for quasar classification.}
    \label{fig:dcpy_feat_imps}
\end{figure*}
\subsubsection{Predicting}
Our approach to classifying a test object involves several steps. First, we
select a distance metric and proceed to scale our data. This scaling is
inspired by the Mahalanobis distance~\citep{mahalanobis1936}: we generalize the
idea of measuring how many standard deviations away a test object is from the
median set for each class calculated in the training step, for the chosen
distance metric. 

For each class, we scale both the test object and the class set by the standard deviation for that class. We then compute the distance between the test object and the median set in units of standard deviations.

Once we have calculated all the distances, we identify the predicted class for the test object as the one for which the object's distance to the median is minimum. 

The detailed prediction algorithm is outlined in \autoref{alg:predict}.

\begin{algorithm}
\caption{Prediction Step}
\label{alg:predict}
\begin{algorithmic}[1]
\STATE Choose a distance metric.
\FOR{each test object}
    \FOR{each class $C$ in the training set}
        \FOR{each feature $F$ in class $C$}
            \STATE Scale the test object feature by dividing it by $\sigma_F^C$, which was calculated in the training step.
            \STATE Scale the median set $\{M_F^C\}$ features by dividing them by $\sigma_F^C$.
        \ENDFOR
        \STATE Calculate the distance $D_C$ between the scaled test object and the median set $\{M_F^C\}$ for class $C$, scaled by $\{\sigma_F^C\}$.
        \STATE Save $D_C$ for class $C$.
    \ENDFOR
    \STATE Choose the class $C_{\text{min}}$ for which the distance $D_{C_{\text{min}}}$ is the smallest.
    \STATE Assign class $C_{\text{min}}$ as the predicted class for the test object.
\ENDFOR
\end{algorithmic}
\end{algorithm}

\subsubsection{Scoring}
\label{subsubsec:scoring}
To evaluate the performance of our classifier's predictions, we use two different scoring methods.
The \fone\ score is given by
\begin{equation} \label{eq:f1_score}
    F_1 = \frac{\text{TP}}{\text{TP} + \frac{1}{2}(\text{FP} + \text{FN})},
\end{equation}
where TP is the number of true positives, FP is the number of false positives, and FN is the number of false negatives. The higher the \fone, the better the classification. 
For more than two classes, we take the average (\emph{macro} mode) of the \fone\ scores for each class. Because we consistently maintain class balance, we do not require any weights when calculating the \fone\ score.

\subsection{Classification Confidence}\label{subsec:confidence}
First, as is standard in the machine learning community, we assess the stability of our
classification results with cross-validation~\citep{CrossValidation95}: we
split our dataset into 5-folds, use four folds as our training set and one fold
as our testing set, and 
average the five \fone\ scores to obtain the final score. This assesses the confidence of the result for a given dataset and the influence of data outliers, but does not provide a direct measure of the uncertainty associated with the use of a specific distance metric.

Therefore, we additionally developed three confidence parameters for use with the classifier. Our goal was to create a distance-based measure for uncertainty (or confidence) in the classification method, not just the data. We tested the performance of the following three distance-derived confidence measures:

\begin{figure}[t!]
    \centering
    \includegraphics[width=0.49\textwidth]{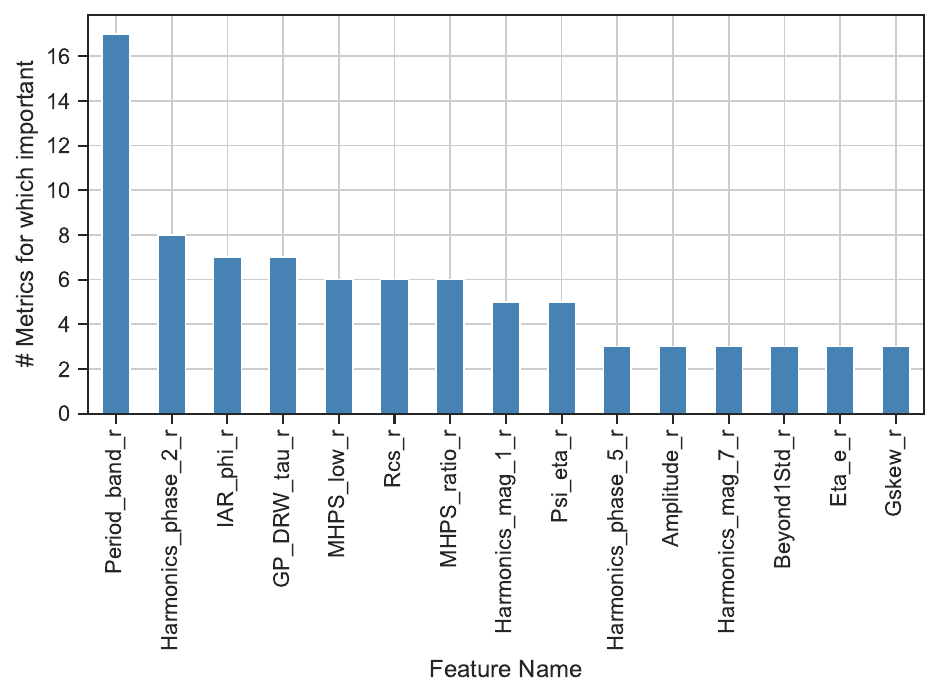}
    \caption{The top features for the multi-class classification task across all metrics. This plot shows 15 features that were included in the final set of $n_{\mathrm{final}}$ features for $\geq$3 metrics. All but one metric (the Maryland Bridge metric) identified \texttt{Period\_band\_r} as important.}
    \label{fig:dcpy_topfeats}
\end{figure}

\begin{figure*}[t!]
    \centering
    \subfloat[Clark Metric]{
        \label{subfig:clark_featimps}
        \includegraphics[width=0.49\textwidth]{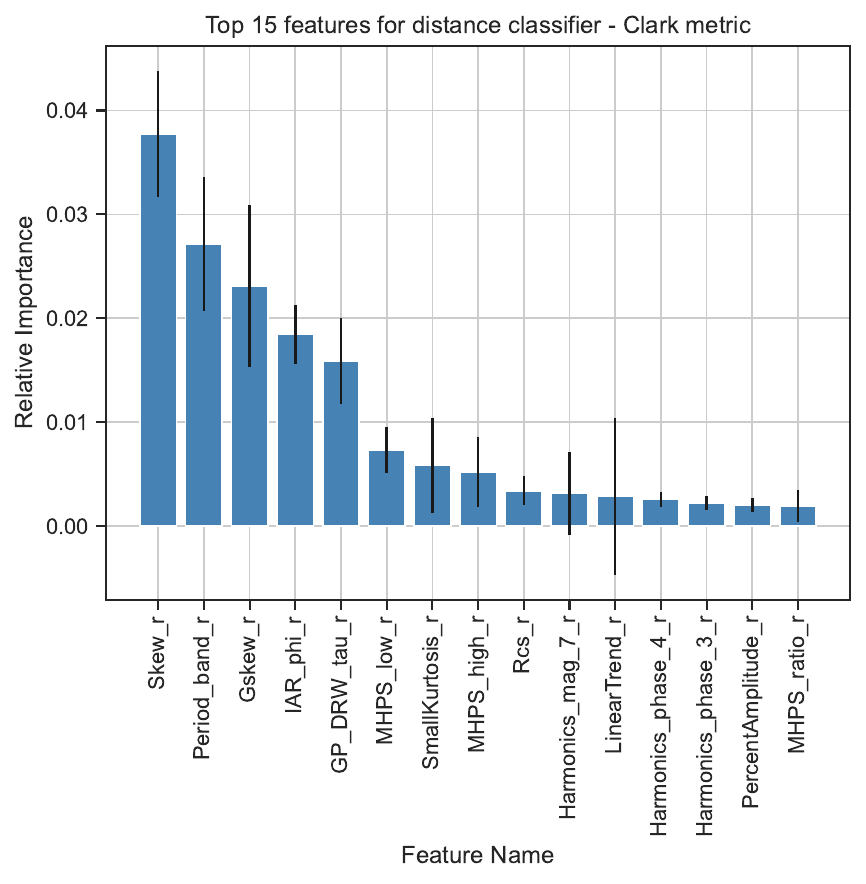}
    }
    \subfloat[Canberra Metric]{
        \label{subfig:canberra_featimps}
        \includegraphics[width=0.49\textwidth]{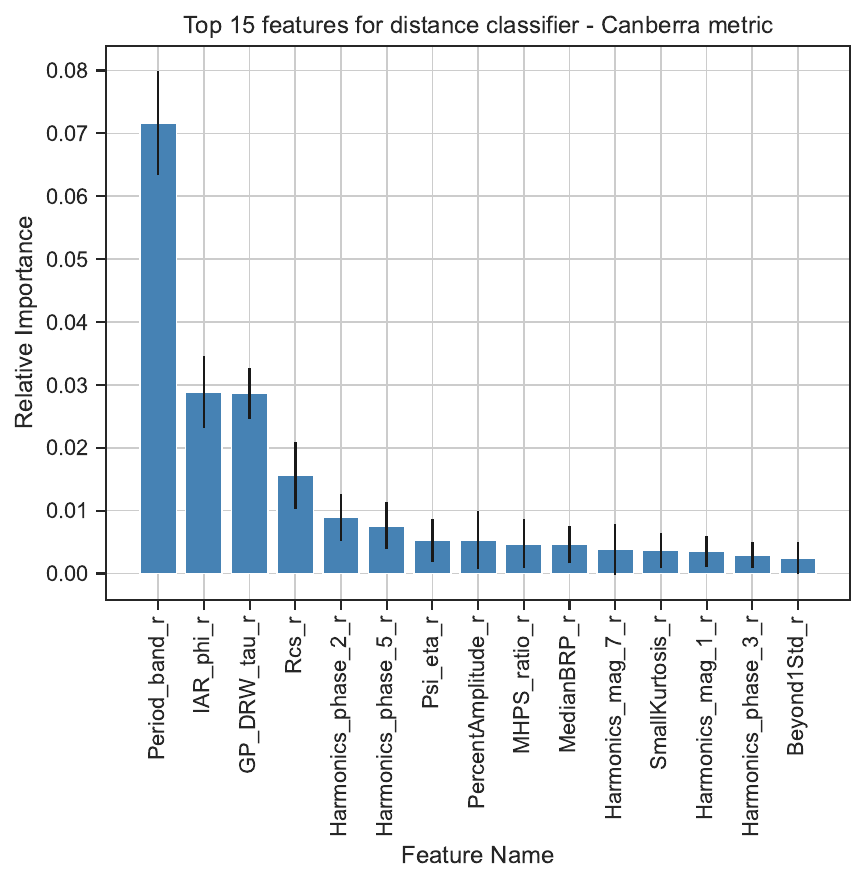}
    }
    \caption{The relative importance of the top 15 features for the multi-class classification task when using the Clark (a) and Canberra metrics (b). The feature importance is determined by computing the Permutation Feature Importance. For each metric, the top 15 features are shown, although only six and three appear significant (at $3\sigma$, respectively). Interestingly, the top feature for the Clark metric is \texttt{Skew\_r} (the skew of the distribution of brightness values in the $r$ band), which is not selected as important with the Canberra metric.}
    \label{fig:dcpy_feature_importances}
\end{figure*}

\begin{itemize}
    
\item Inverse of total distance: 
We invert the distance ($d$) calculated in the prediction step to give a classification confidence, such that a low distance value corresponds to a higher confidence value. Note that this distance is calculated in $n$ dimensions corresponding to the $n$ features, and depends on the metric being used. Furthermore, it has already been scaled by the standard deviation during computation (\autoref{alg:predict}). Since the distance scales are different for different metrics, this value is only useful for comparison of confidences among different objects for a given metric. However, for a given metric, this is very easily interpretable, as something having a zero distance would imply an almost infinite confidence.
\[
 c_d = \dfrac{1}{d} 
\]
\item Inverse of scaled distances: For a given feature ($i$), we compute a one-dimensional distance ($d_i$) between the test object and the median set. Because each feature has a different level of variance, we then scale it with the standard deviation ($\sigma_i$) associated with that feature within the reference class, and then finally invert it.
\[
 c_{\text{scaled}} = \dfrac{1}{\Sigma_i \frac{d_i}{\sigma_i}} 
\]
Scaling the distances with feature-specific standard deviations, we can ensure that the resulting confidence value can be compared across different metrics. However, this distance is not applicable for all metrics because not all distance metrics are defined in one dimension.
    
\item Kernel Density Estimate ($c_{\text{KDE}}$) Probability: Kernel Density
Estimate (KDE) is a statistical technique to estimate the
probability density from given
data~\citep{silvermanDensityEstimationStatistics2018}. It is
given by:
\[
f_{\text{KDE}}(x) = \frac{1}{Nh} \sum_{i=1}^{N} K\left(\frac{x - x_i}{h}\right),\]
where $x$ is the continuous variable over which density is estimated, $N$ is the number of data points, $h$ is the bandwidth parameter, $x_i$ are the points in the dataset and $K$ is the kernel function (for which we choose a Gaussian).

We thus use this KDE to generate the probability density for each class given its features. Finally, to calculate the KDE probability for a test object ($x_{\text{test}}$), we plug its features into the KDE for that class, and get the confidence for a test object belonging to that class ($c_{\text{KDE}}$)
\[
c_{\text{KDE}} = f_{\text{KDE}}(x_{\text{test}})
\]
\end{itemize}

\begin{figure}[bh!]
    \centering
    \includegraphics[width=0.49\textwidth]{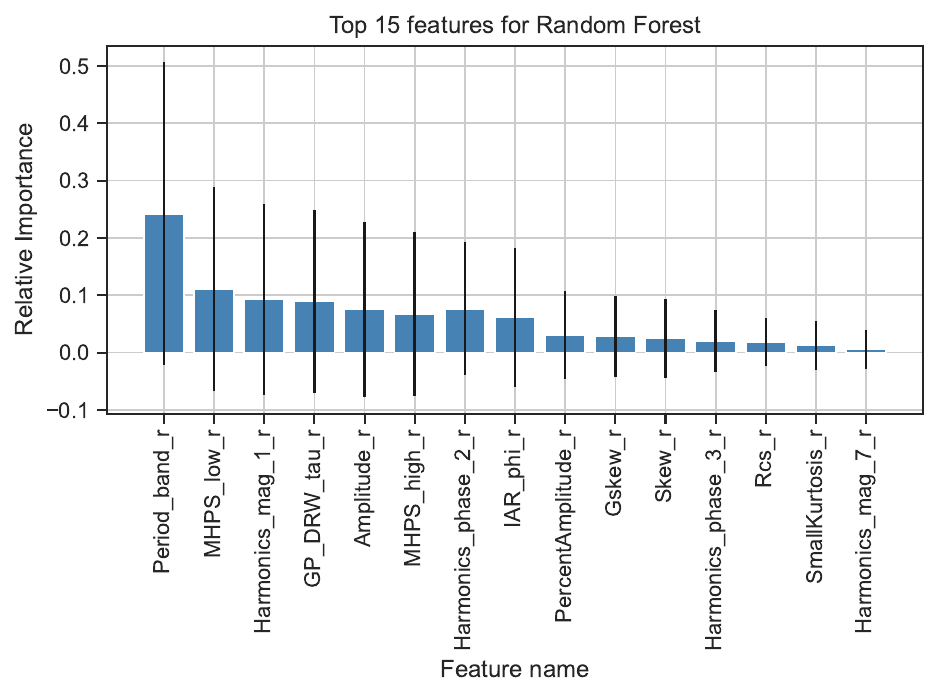}
    \caption{The top 15 features for a Random Forest Classifier for the multi-class classification task mentioned in \autoref{subsec:classification_problems}. The feature importance is determined by computing the Gini Importance~\citep{breimanClassificationRegressionTrees1984}. We note that none of the features' importance is statistically significant.}
    \label{fig:feature_importance_rf}
\end{figure}

\begin{figure*}
    \centering
    \subfloat[Clark Distance Classifier ($F_1 = 92.0 \%$)]{
        \label{subfig:clark_cm}
        \includegraphics[width=0.33\textwidth]{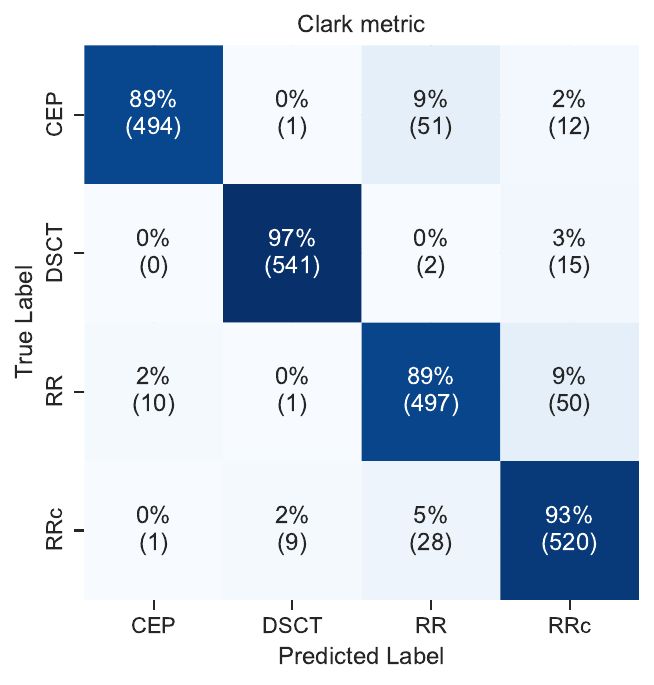}
    }
    \subfloat[Canberra Distance Classifier ($F_1 = 91.2 \%$)]{
        \label{subfig:canberra_cm}
        \includegraphics[width=0.33\textwidth]{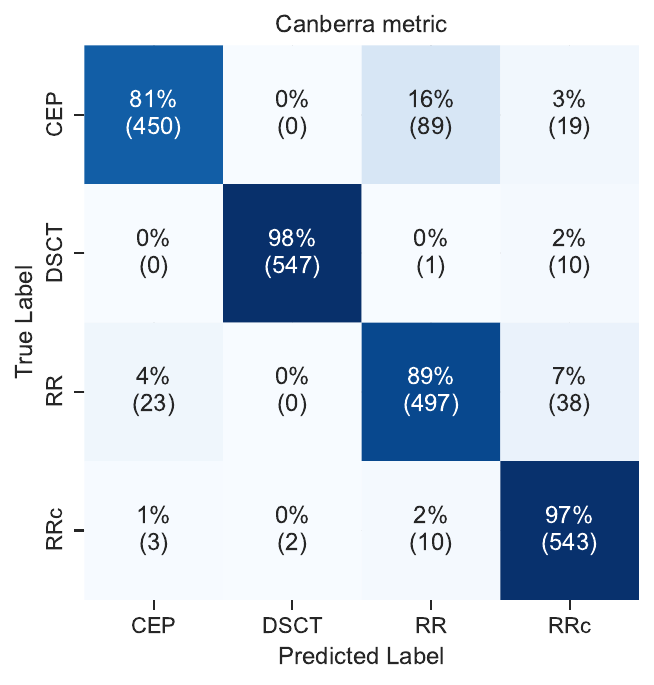}
    }
    \subfloat[Random Forest Classifier ($F_1 = 92.0 \%$)]{
        \label{subfig:rf_cm}
        \includegraphics[width=0.33\textwidth]{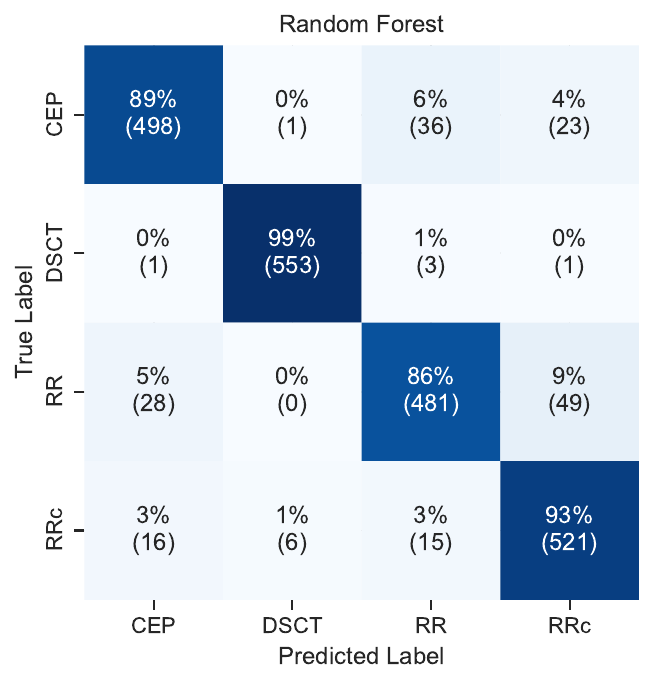}
    }

    \caption{Cross-Validation Results: Confusion matrix of our \distclassipy\ model using Clark (\emph{left}) and Canberra (\emph{center}) distances (two of the top-performing distance metrics) compared to the performance of a  Random Forest Classifier (\emph{right}) for the multi-class classification task of Cepheids (CEP), Delta Scuti (DSCT), RR Lyrae ab (RR) and RR Lyrae c (RRc). The respective \fone\ scores are shown in each subplot: all three models achieve comparable performance ($F_1\sim92\%$). Each cell in each plot shows the percentage and the number of objects corresponding to that combination of label ($y$) and prediction ($x$). The diagonal corresponds to correct classifications.}
    \label{fig:confusion_matrices}
\end{figure*}

\subsection{Sequential Feature Selection and Classification}
We have already reduced our feature space from 114 to 31 features. To increase the robustness of our method and decrease the computational cost, we further reduce our dimensionality by selecting the most effective classification features for each class and each problem. This process is computationally expensive (as described below) and the final set of features to be used depends on the metric used and the classification task. However, this computation can be performed once for each problem, as we did for the three classifications we explored in our work. Thereafter, the user of our algorithm can choose the features that best match their research based on their scientific interest and object sample, with a net computational gain. Here, we describe our strategy, and the detailed results are discussed in \autoref{subsec:results_feat_imp}.

We use the Forward Sequential Feature Selection (SFS)
strategy \citep{ferriComparativeStudyTechniques1994}. Forward SFS works by
adding one feature at a time: we start by choosing the single feature that,
alone, maximizes our 5-fold cross-validation \fone{} score. We continue adding
features and increasing the dimensionality of the feature space, each time
selecting the next feature that maximizes \fone. We use Forward SFS over
Backward SFS because of the high computational requirements of Backward SFS
which make it prohibitive in a high dimensional space. Recursive Feature
Elimination (RFE) is another option for the selection of features, but RFE
requires a built-in feature importance attribute to the model, which is not
implemented at this time in \distclassipy. A comparison of the selection
performed by RFE, compared to SFS, is left for future work. 

The total number of possible feature combinations are,

\begin{align}
    \text{No. of combinations} &= \sum_{i=0}^{n} \binom{n-i}{1} \notag \\
    &=\frac{n(n+1)}{2} \label{eq:tot_sfs_calcs}\\
    &=\mathcal{O}(n^2) \label{eq:bigo_sfs}
\end{align}

With $n=31$, we perform a total of $465$ calculations for each distance metric (18) and classification task (3). 
This provides us with the feature importance ranking and
 allows us to track how performance varies with the number of included features, enabling us to 
optimize the dimensionality of our feature space: in each case, we choose a final set of $n_{\mathrm{final}}$ features such that $n_{\mathrm{final}}$ is the lowest number of features whose \fone\ score is within $1\sigma$ of the maximum \fone\ score. An example of SFS is shown in \autoref{fig:sfs_plots}, for the Canberra and Clark distance metrics in the multi-class classification case.

The feature importance is measured as the relative permutation feature
importance, an algorithm-agnostic method~\citep[originally introduced in random forests;][]{breimanRandomForests2001} where the importance of a feature is
measured as the drop in performance when the feature is randomly shuffled
during testing. Feature importance for the multi-class classification is shown
in \autoref{fig:dcpy_feat_imps} for each distance metric.

In \autoref{fig:dcpy_topfeats}  we show the number of distance metrics (out of our 18 metrics) for which a given feature is selected. The 15 features shown were important for $\geq$3 metrics and are considered as a ``super-set'' of important features. We emphasize that the user should select the most effective distance metric and subset of features, performing a feature selection such as the one described above, specifically for their data and classification problem, as the classification goal (\eg,  purity \textit{vs} completeness in selecting a specific class of objects, \textit{vs} separating two specific subclasses, etc.), as well as data quality (including sparsity, signal-to-noise ratio, noise property, number of bands, etc.) will influence the selection of features and ultimate performance of the model.

The top features for the Clark and the Canberra metrics are shown in \autoref{fig:dcpy_feature_importances} for the multi-class classification case. 
For comparison, we also show the feature importance results of the RFC in \autoref{fig:feature_importance_rf}. We note that none of the RFC features' importance reaches statistical significance.

\begin{table*}
\centering
\renewcommand{\arraystretch}{1.15} 
\begin{tabular}{ll|cc|cc|cc|}
\cline{3-8}
\multicolumn{2}{l|}{}                                                           & \multicolumn{2}{c|}{\textbf{Multi-class}} & \multicolumn{2}{c|}{\textbf{Binary}} & \multicolumn{2}{c|}{\textbf{One-vs-rest}} \\ \cline{3-8} 
                                                                &               & \distclassipy                & RFC                & \distclassipy              & RFC              & \distclassipy                 & RFC                \\ \hline
\multicolumn{1}{|l|}{\multirow{5}{*}{\textbf{\fone{} Score (\%)}}}   & Max           & 92                   & 92                 & 68                & 70               & 94                   & 95                 \\
\multicolumn{1}{|l|}{}                                          & 75 Percentile & 91                   &                    & 66                &                  & 93                   &                    \\
\multicolumn{1}{|l|}{}                                          & 50 Percentile & 90                   &                    & 65                &                  & 93                   &                    \\
\multicolumn{1}{|l|}{}                                          & 25 Percentile & 82                   &                    & 64                &                  & 90                   &                    \\
\multicolumn{1}{|l|}{}                                          & Min           & 75                   &                    & 62                &                  & 89                   &                    \\ \hline
\multicolumn{1}{|l|}{\multirow{5}{*}{\textbf{No. of features}}} & Max           & 20                   &                    & 25                &                  & 17                   &                    \\
\multicolumn{1}{|l|}{}                                          & 75 Percentile & 7                    &                    & 3                 &                  & 5                   &                    \\
\multicolumn{1}{|l|}{}                                          & 50 Percentile & 4                    &                    & 3                 &                  & 4                    &                    \\
\multicolumn{1}{|l|}{}                                          & 25 Percentile & 4                    &                    & 2                 &                  & 2                    &                    \\
\multicolumn{1}{|l|}{}                                          & Min           & 1                    &                    & 1                 &                  & 1                    &                    \\ \hline
\end{tabular}
\caption{Summary table of the performance of \distclassipy\ for each classification task. The statistics are reported across the selection of distance metrics and feature subsets. The RFC results are shown for comparison next to the highest \dmc\ performance for each classification task. The complete set of results is available online as a spreadsheet {\href{https://docs.google.com/spreadsheets/d/1cNaXAjW_RMu3y6MUPkNmjOs03kbRjExU9YLIAjg8A-M/edit?usp=sharing}{here}}.}
\label{tab:result_summary}
\end{table*}

\subsection{Random Forest Classifier}
We use a Random Forest Classifier~\citep[RFC, ][]{breimanRandomForests2001} as
a benchmark to compare our method. A Random Forest involves training an
ensemble of decision trees, each performing binary decisions on one feature at
a time. This method, therefore, does not leverage or require the definition of
a distance. The RFC was run in the  \texttt{scikit-learn} implementation with
the following hyperparameter choices: the maximum depth was set to 3 to avoid
overfitting, the forest was composed of 100 estimators (trees), with
bootstrapped samples (to reduce the variance) and considering $\sqrt{N}$
features at each split. As for the distance-based models, the RFC was run with
a 5-fold cross-validation scheme on the entire dataset.

\section{Results} \label{sec:results}

In this section, we report and discuss the performance of our model (\distclassipy) and the comparison model (RFC), as well as the results of our investigation into the classification's confidence and feature selection. 

\subsection{\distclassipy\ classification performance}
\label{subsec:results_performance}

The best cross-validation result of our \dmc\ (DMC) from \distclassipy\ across all distance metrics and feature subsets for each classification task are as follows: for the Multi-class Classification, $\text{DMC}=92\%\ \text{(}\text{RFC}=92\%$); for the \ovr, $\text{DMC}=94\%\ \text{(}\text{RFC}=95\%$; for the Binary classification of rotating stars, $\text{DMC}=68\%\ \text{(}\text{RFC}=70\%$. A comprehensive report of the performance of our model, including the distribution of performance values across different distance metric and feature selection choices, is presented in \autoref{tab:result_summary}. The full spreadsheet is available online.\footnote{\href{https://docs.google.com/spreadsheets/d/1cNaXAjW_RMu3y6MUPkNmjOs03kbRjExU9YLIAjg8A-M/edit?usp=sharing}{Cross-Val Results URL}} 
Confusion matrices for the multi-class classification task for the \dmc\ (using the top two distance metrics: Clark and Canberra) and RFC are shown in \autoref{fig:confusion_matrices}. From these, we see that performance is mostly very similar. However, for some classes like RR and RRc, the Canberra distance classifier outperforms the RFC. 

Finally, we test our final classifiers on the hidden set introduced in \autoref{subsec:catalog}, which consists of $500$ objects, entirely unseen in training. For each classification, we test the best model (best distance metric and features subset) on a hidden dataset, first rebalanced and then unbalanced,  where each class has a representation similar to the fraction of objects in the initial dataset. We confirmed the model performance for all three classification tasks, and found the results to be mostly within $1 \sigma$ expectations from our cross-validation results for the rebalanced dataset. With the best distance metric selected on the cross-validated data, for each classification task the performance on the hidden set is $F_1 = 93\%$ with the Clark metric on the multi-class classification (1881 objects), $F_1 = 68\%$ with the Euclidean metric on the binary classification (1000 objects) and $F_1=92\%$ with the Kulczynski metric for the \ovr\ classification (976 objects; \autoref{sec:classification}). For the unbalanced dataset, the performances change slightly. In the multi-class classification, the unbalanced dataset contains 494 RR, 253 DSCT, 210 RRc, 24 CEP. The performance of the Canberra-based model improves slightly from $F_1 = 91\%$ to $F_1 = 93\%$, while the performance of the Clark-based model drops to $F_1 = 86\%$ from 92\%. This is primarily due to confusion between the RR and RRc classes, which are the most common labels. However, for comparison, the RFC performance drops significantly more, from $F_1 = 92\%$ to $F_1 = 76\%$.  In the imbalanced \ovr\ classification, the performance drops by a few points, from $F_1 = 93\%$ (with Euclidean distance) to $F_1 = 90\%$. Similarly, the RFC performance drops from $F_1 = 95\%$ to $F_1 = 89\%$. The binary classification is naturally balanced with 500 BYDra and 480 RSCVN.

The reader should however be reminded that our classification is based on the
labels found in~\citet{chenZwickyTransientFacility2020a} and that
while~\citet{chenZwickyTransientFacility2020a} use DR2 ZTF data, our work is
based on data from DR15. A possible implication, since DR2 data contains fewer
points per light curve, is that label noise contributes to apparent inaccuracy
in our scores.

\begin{figure*}
    \centering
    \subfloat[Clark Distance Classifier on the \emph{hidden} set ($F_1 = 92.7 \%$)]{
        \label{subfig:clark_cm_hidden}
        \includegraphics[width=0.4\textwidth]{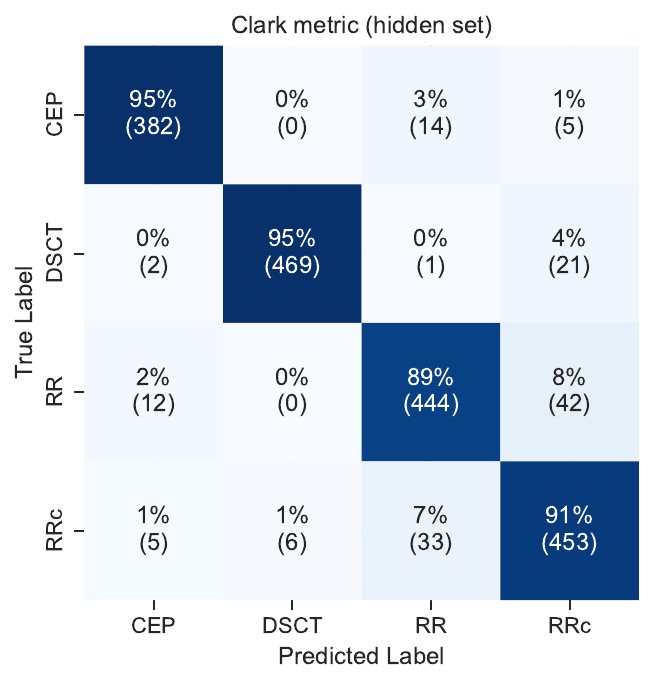}
    }
    \hspace{4em}
    \subfloat[Canberra Distance Classifier on the \emph{hidden} set ($F_1 = 94.7 \%$)]{
        \label{subfig:canberra_cm_hidden}
        \includegraphics[width=0.4\textwidth]{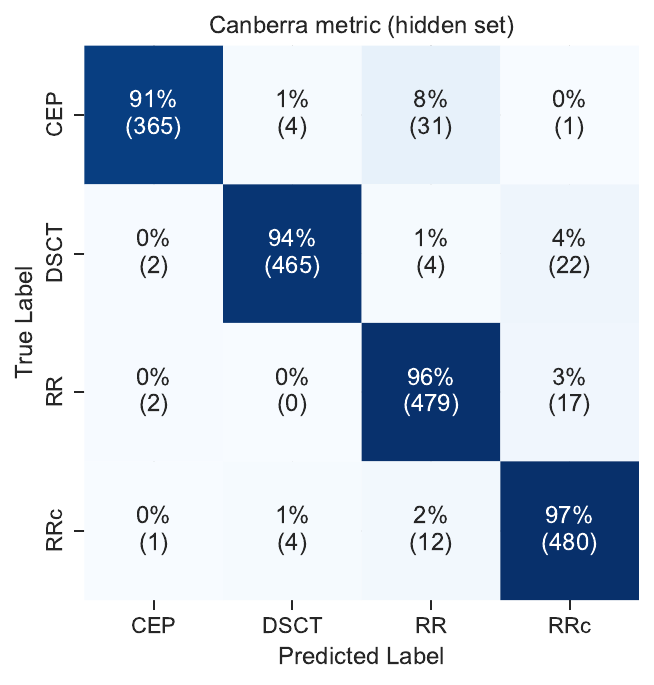}
    }

    \caption{Hidden Set Results: Confusion matrices (as in \autoref{fig:confusion_matrices}) for our \distclassipy\ model on the \emph{hidden} set using Clark (\emph{left}) and Canberra (\emph{right}) distances for the multi-class classification task of Cepheids (CEP), Delta Scuti (DSCT), RR Lyrae ab (RR), and RR Lyrae c (RRc). Both Clark and Canberra perform more than 1 $\sigma$ better than their scores from cross-validation ($91.99\% \pm 0.61\%$ and $91.22\% \pm 1.38\%$, respectively).}
    \label{fig:cm_hidden}
\end{figure*}

The complete set of results is available online as a spreadsheet.\footnote{\href{https://docs.google.com/spreadsheets/d/1VJuiFfFPaHRqjBkwyAbPre3rxUZ04PZMaI5SJWC7QNE/edit?usp=sharing}[Hidden Set Results URL]}. The confusion matrices for the multi-class classification of the hidden set are displayed in \autoref{fig:cm_hidden} for the Clark and Canberra metric.

For the remainder of the discussion, we will focus solely on the multi-class classification task.

\subsection{Computational Requirements}
\label{subsec:results_timing}
\begin{figure*}
    \centering
    \subfloat[Visual comparison of computation time across a Random Forest, and \distclassipy\ with Clark and Canberra metric.]{
        \label{subfig:timing_circles}
        \includegraphics[width=0.7\textwidth]{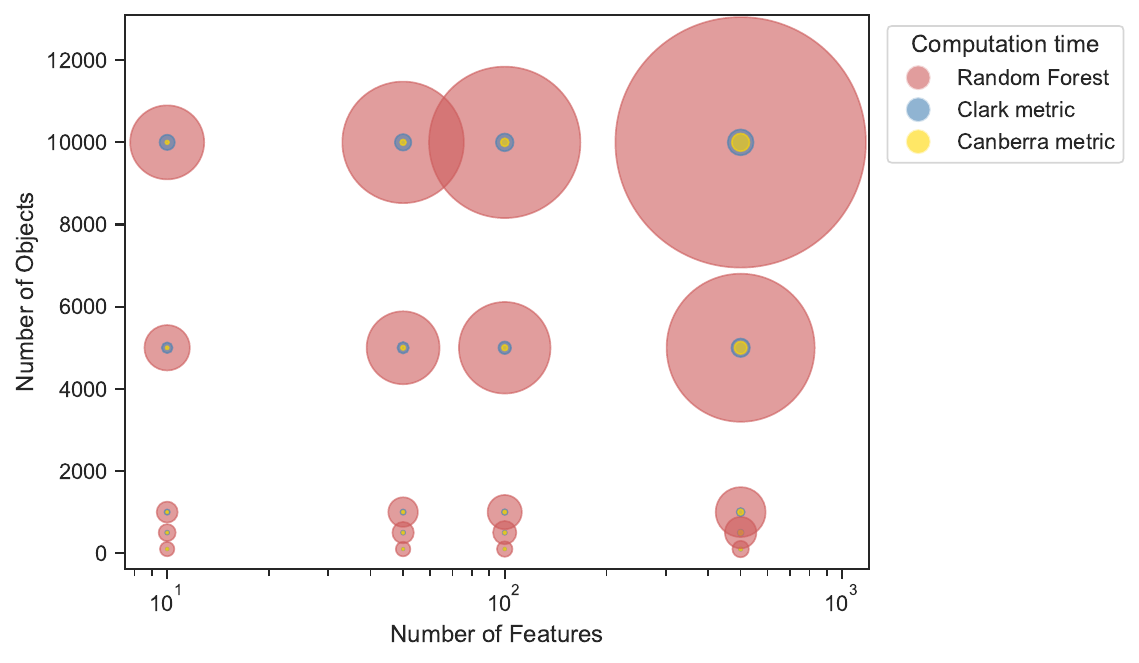}
    } \\
    \subfloat[Clark Distance Classifier]{
        \label{subfig:timing_clark}
        \includegraphics[width=0.32\textwidth]{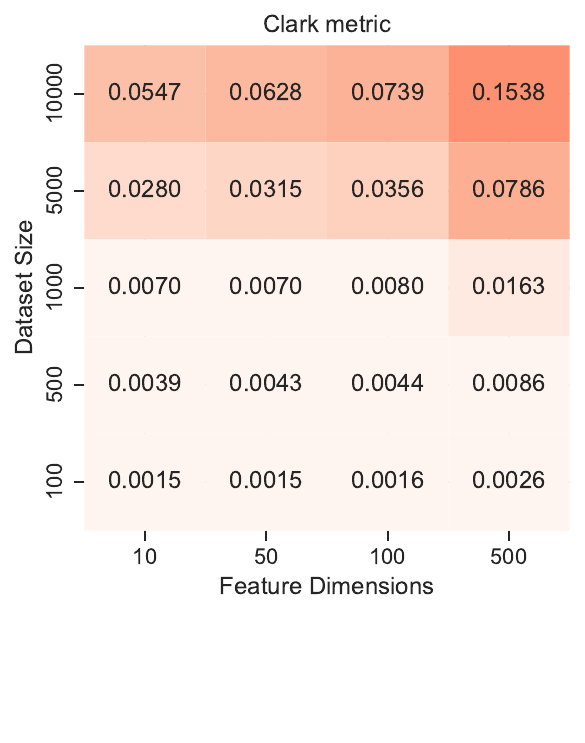}
    }
    \subfloat[Canberra Distance Classifier]{
        \label{subfig:timing_canberra}
        \includegraphics[width=0.32\textwidth]{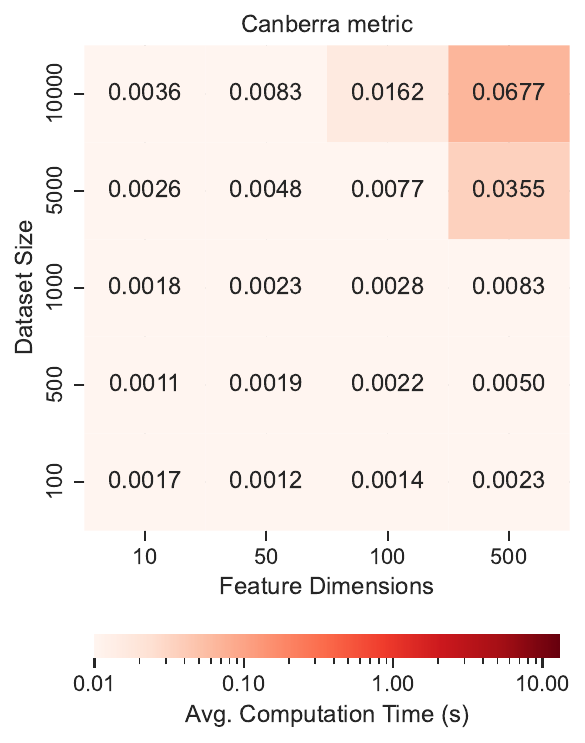}
    }
    \subfloat[Random Forest Classifier]{
        \label{subfig:timing_rfc}
        \includegraphics[width=0.32\textwidth]{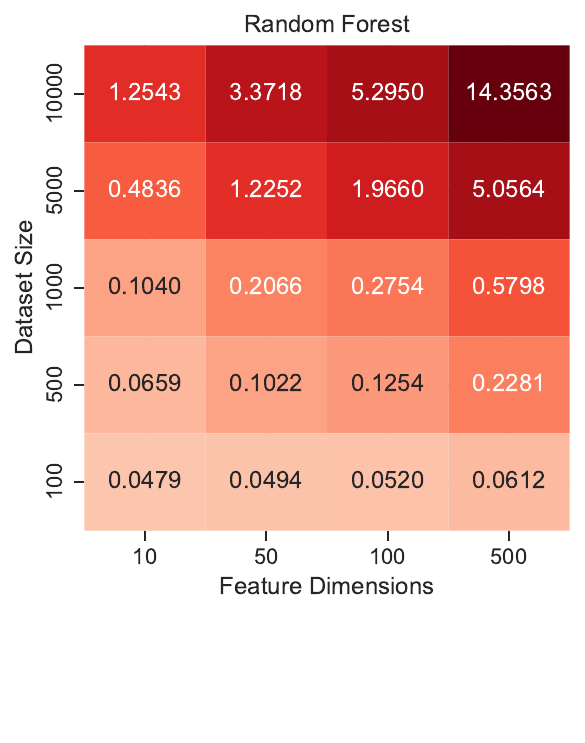}
    }
    \caption{Top: A visual comparison of computational times between our method, \distclassipy\ and RFC as a function of the dataset size (number of objects; $y$-axis) and number of features ($x$-axis). For all plots, the computational costs of fitting the model on the training data and predicting on the testing data (not including preprocessing) are evaluated on the entire dataset five times, and the results shown are the average over these five iterations. The area of each circle is proportional to the computational time required for training and generating the predictions. We see that \distclassipy\ is always faster than RFC (with 100 trees), and the difference is even more prominent in larger datasets.  Bottom: A comparison of computational times (in seconds) between \distclassipy\ and RFC as a function of variable dataset size and number of features. We see that \distclassipy\ is always at least an order of magnitude faster than RFC (with 100 trees), even when using custom metrics (\eg\ Clark) which are not in-built in \texttt{SciPy} (like, \eg, Canberra).}
    \label{fig:timing}
\end{figure*}

\begin{figure*}[t!]
    \centering
    \subfloat{
        \includegraphics[width=0.25\linewidth]{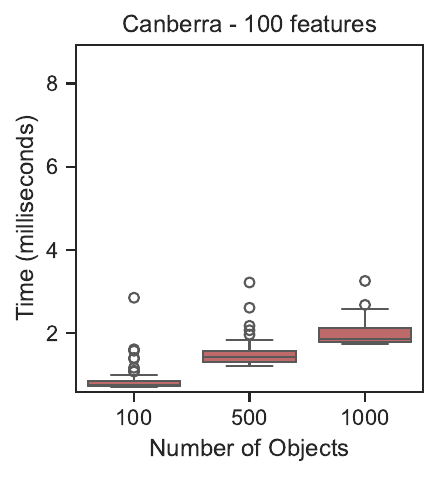}
        }
    \subfloat{
        \includegraphics[width=0.25\linewidth]{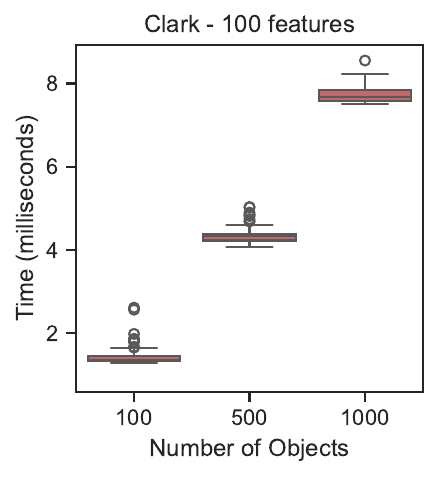}
        }\\
    
    \subfloat{
        \includegraphics[width=0.25\linewidth]{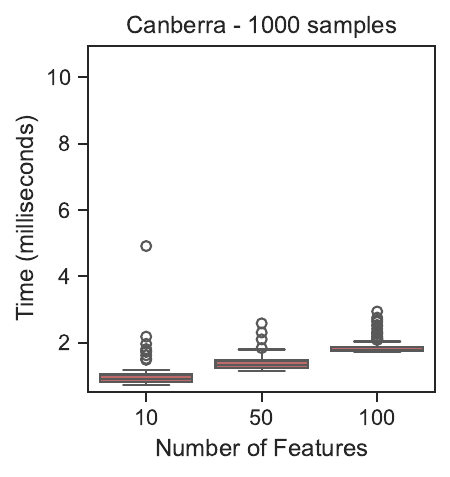}
       }
    \subfloat{
        \includegraphics[width=0.25\linewidth]{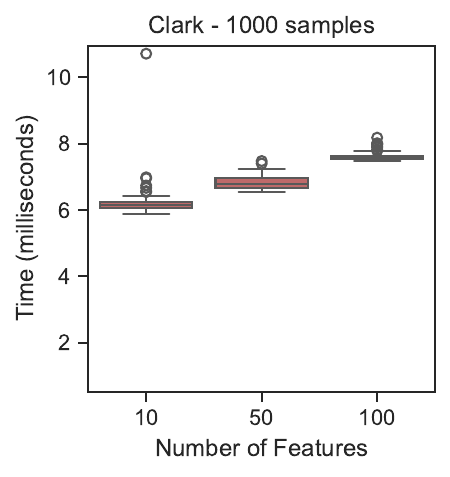}
       }
    \caption{The computational time of \distclassipy\ using the Canberra and Clark metrics (left and right columns respectively) as a function of sample size using 100 features (top plots) and of number of features using 1000 objects (bottom plots).}
    \label{fig:computationaltime}
\end{figure*}

To understand how the computational time scales with dataset size and feature dimensions, we monitored the total time taken for training and prediction for different numbers of data points and features using synthetic data generated with \texttt{scikit-learn} --- the specifications for our hardware and software are outlined in \ref{app:specs}. We compared the timings of \distclassipy\ using the metrics Canberra and Clark. It is worth noting that the Canberra metric is implemented in the \texttt{scipy.spatial.distance} package, and we thus expect it to be computationally optimized. On the other hand, the Clark metric is not available within \texttt{Scipy} (to our knowledge) and we coded it ourselves: we expect it may be sub-optimal in terms of computational costs. The timings of an RFC with 100 trees is also tracked. We ran each method five times, and report the mean here. Our results are illustrated in \autoref{fig:timing} and \autoref{fig:computationaltime}.

We find that \distclassipy\ is remarkably scalable and shows minimal computation times even for large and high-dimensional datasets. Even when dealing with a dataset size of 10,000 objects and 500 features, the computation times for \distclassipy\ remain as low as $0.0677 \pm 0.0016$ seconds (Canberra) and $0.1538 \pm 0.0023$ seconds (Clark) compared to $14.3563 \pm 0.4978$ seconds for RFC. These make \distclassipy\ very suitable for current and future astrophysical surveys, with increasing sample sizes (\eg, the Vera C. Rubin LSST is expected to characterize 17B stars, many of which are variable to the depth limit of the survey, and 48B astrophysical objects altogether).

Finally, \distclassipy\ and RFC utilize a comparable amount of RAM (the \dmc\ utilized $574.5 \pm 4.8 $ MiB compared to RFC's $568.4 \pm 16.3 $ MiB on our machine.\footnote{Note: These numbers are meant for comparison when run on our machine whose specifications are given in \ref{app:specs}} for the dataset having 10,000 objects and 500 features) during training, as per our tests using the Python package \texttt{memory\_profiler}.

In conclusion, \distclassipy\ performs comparably to the RFC (state-of-the-art), while being faster and utilizing a comparable amount of memory. 

We now look at other assets that \distclassipy\ can offer us to help understand classification --- confidence and robustness.

\subsection{Feature Importance}
\label{subsec:results_feat_imp}
\begin{figure}
    \centering
    \includegraphics[width=\columnwidth]{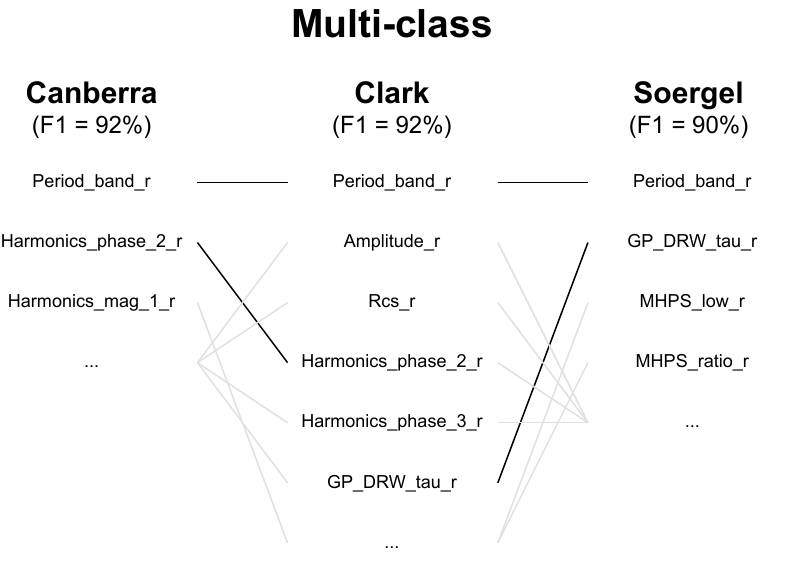}
    \caption{
    How feature importance changes for three different distance metrics (Canberra, Clark, and Soergel) in the multi-class classification. Each of these metrics has a similar performance, but there is only one feature that is common among all of them: \texttt{Period\_band\_r} --- the $r$-band period.}
    \label{fig:feat_imp_flow}
\end{figure}

We find that whether a feature is important or not depends not only on the classification task, but also the distance metric being used. In other words, each metric has features with which it works best, and often, these features differ for different metrics. We show in \autoref{fig:feat_imp_flow} how the important features change for the multi-class classification task for three metrics --- Canberra, Clark, and Soergel.

This flexibility allows for the \dmc\ to be adapted to the specific research goal and resources. In a scenario for a particular classification task, if some features are easier or faster to calculate than others, one could use \distclassipy\ with a distance metric which is the most optimal for these features.

\subsection{Confidence of the distance metric classification}
\label{subsec:results_confidence}
\begin{figure}
    \centering
    \includegraphics[width=0.49\textwidth]{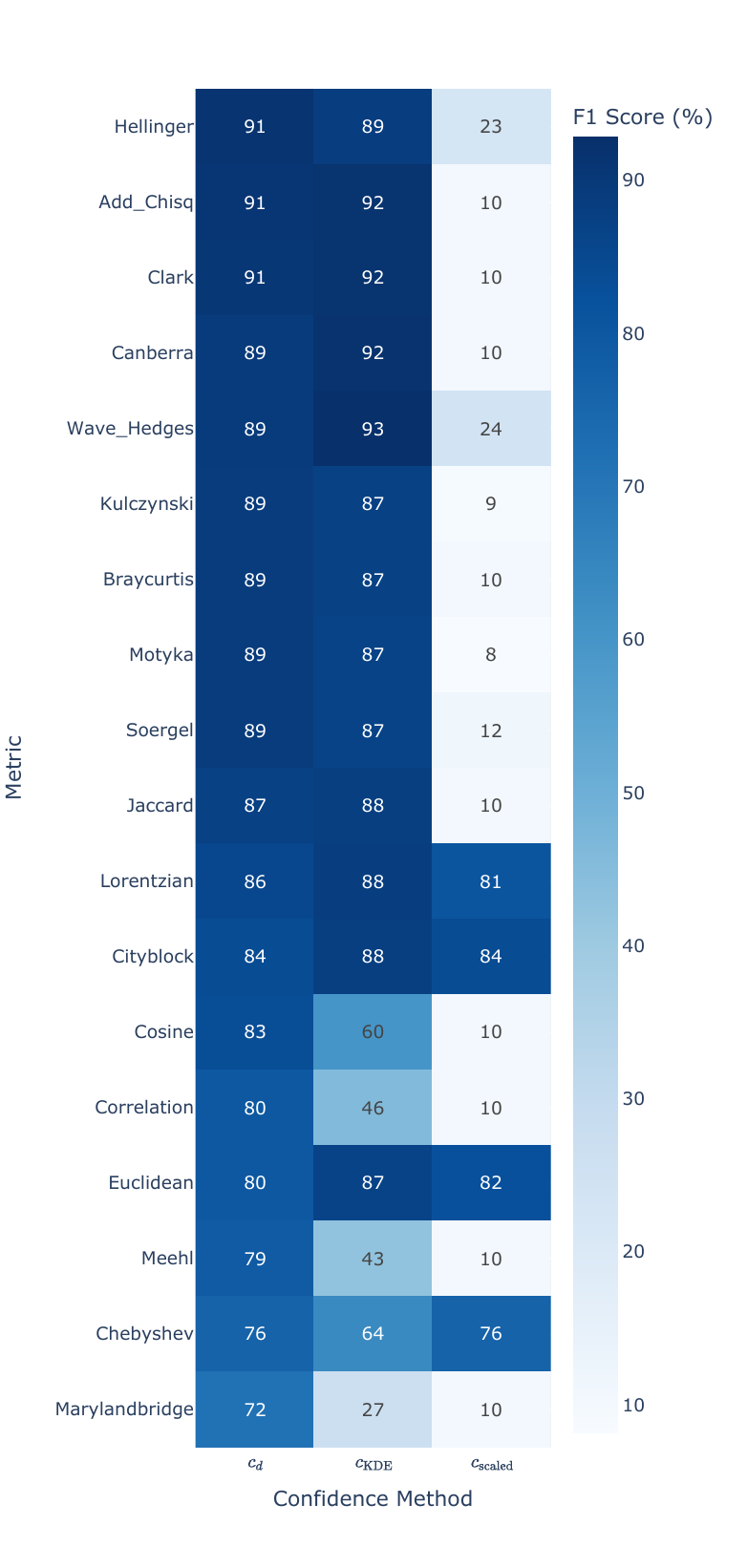}
    \caption{Here, we compare the effectiveness of three confidence methods mentioned in \autoref{subsubsec:scoring} for our 18 metrics, for the multi-class classification task. We do this by treating our confidences as a pseudo-probability, and then using this pseudo-probability for predicting a test object. Finally, we calculate the accuracy for each method using these predictions, which has been plotted above. This accuracy score helps tell whether a confident prediction is an accurate prediction. From this plot, we see that $c_{\text{scaled}}$ and $c_{\text{KDE}}$ are ineffective methods, and we thus choose $c_d$ as the confidence method for \distclassipy.}
    \label{fig:dcpy_metrics_heat}
\end{figure}

In \autoref{subsec:confidence}, we introduced three different confidence methods with the goal of measuring the classification reliability for each metric. Here, we compare these three methods. The question we ask is - does a confident prediction imply a correct prediction?
Although it is difficult to answer this in absolute, we can instead employ the following strategy: we treat the confidence as a pseudo-probability, and then see what the performance would be if we used this probability as a class prediction (instead of the \dmc). Finally, we evaluate the \fone\ score for the prediction for each of the three confidence methods \autoref{fig:dcpy_metrics_heat}. We note that, because the first confidence method, $c_d$ is just $1/d$, where $d$ is the \dmc\ distance, this prediction is the same as the \distclassipy\ prediction.

The \fone\ scores for predictions from the method $c_d$ is the most reliable, and works well for all distance metrics ($ 72\% \leq F_{1,c_d} \leq 91\% $). $c_{\text{KDE}}$ slightly outperforms the $c_d$ method for a few metrics (\eg, Wave-Hedges), but dramatically fails for some others (\eg, Maryland-bridge). However, all of the metrics for which $c_{\text{KDE}}$ underperforms ($\ F_{1,c_{\text{KDE}}} \leq 64\% $) are the lowest five performers overall, while for all others $ F_{1,c_{\text{KDE}}} \geq 87\% $. Meanwhile, scores for $c_{\text{scaled}}$ are the poorest for all distance metrics. Based on this we conclude that, although the simplest, the confidence parameter $c_d$ is a more reliable indicator of overall classification confidence and we thus use $c_d$ in \distclassipy. 

An interesting point to note is that, unlike the probabilistic outputs from something like RFC, our \dmc\ with $c_d$ does not assume a fixed confidence level shared across all classes. The \dmc\ does not impose a strict boundary on how far away a test object can be, and we suspect this may be useful for dealing with outliers and anomalies. However, this is not something we will explore in this work, and we end our discussion on confidences. 

\subsection{Robustness of the distance metric classification}
\label{subsec:results_robustness}

\begin{figure*}
    \centering
    \subfloat[Clark metric]{
        \label{subfig:clark_robustness}
        \includegraphics[width=0.49\textwidth]{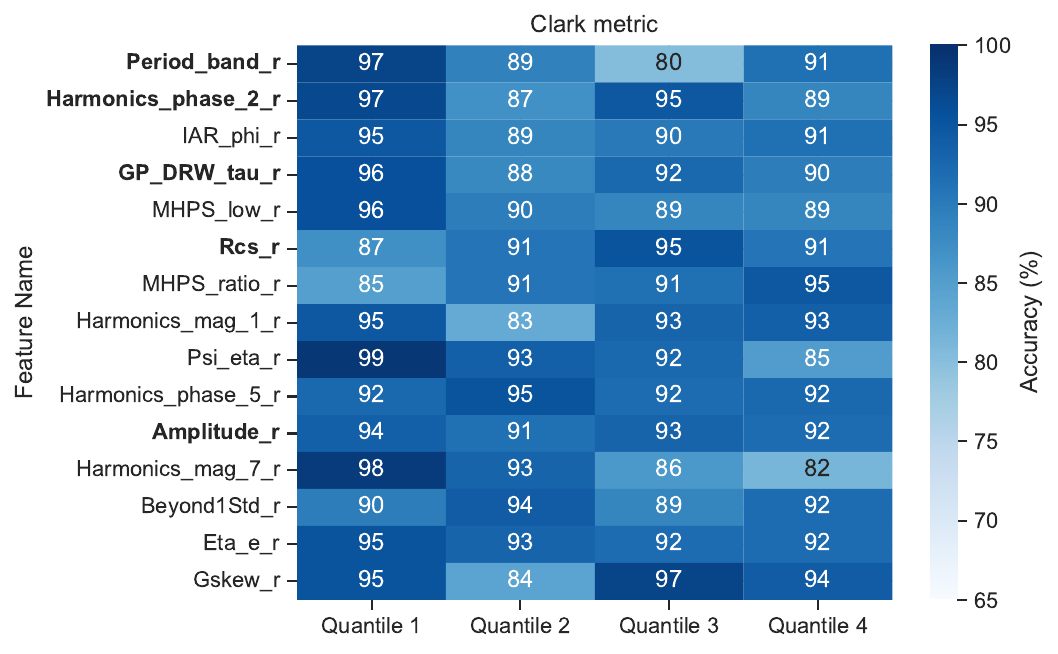}
    }
    \subfloat[Canberra metric]{
        \label{subfig:canberra_robustness}
        \includegraphics[width=0.49\textwidth]{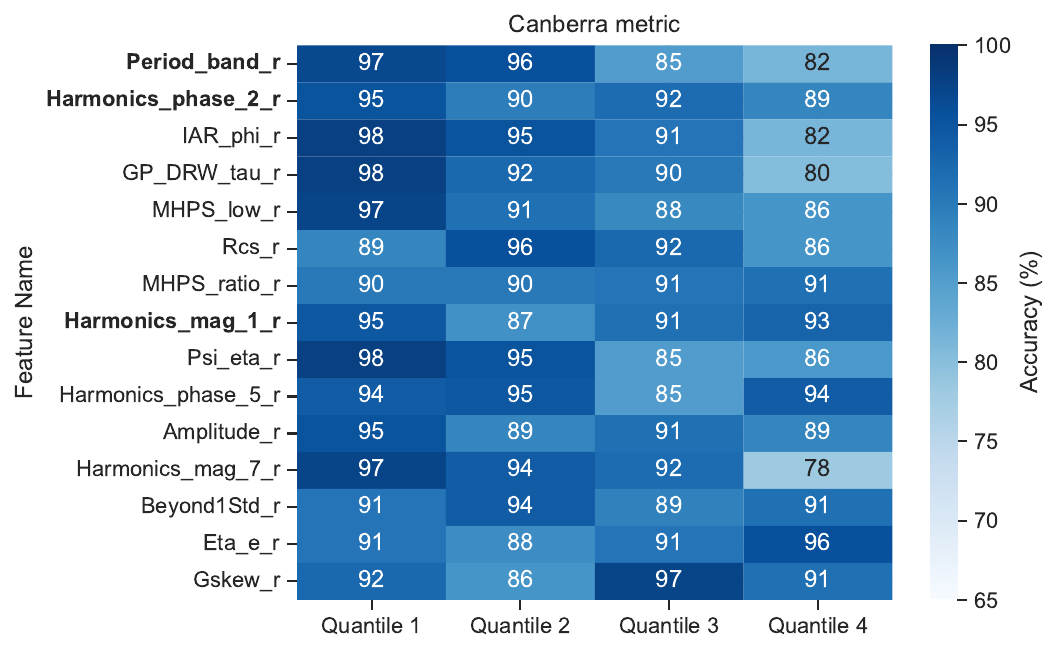}
    }
    \caption{Robustness scores for the multi-class classification task, with the \dmc\ using the Clark (\autoref{subfig:clark_robustness}) and Canberra metrics (\autoref{subfig:canberra_robustness}). The robustness score is calculated by splitting our test set into quantiles, and then evaluating the classification accuracy for each quantile. Note that, although each quantile is used to split the dataset, only a subset of these features are actually used in the classification (boldfaced). We still include all features from the super-set of features for the quantile split, to better compare performance for different metrics. For \eg, although the Canberra metric (b) performs better in general, Clark metric (a) performs better with objects having larger periods (quantile 4, top row). The color maps limits for \autoref{fig:robustness_plots} and \autoref{fig:robustness_bestfeat} have been cut to the range [65,100] to increase the readability of differences in performance, but both models presented in this figure perform better than 78\% at any quantile with any feature.}
    \label{fig:robustness_plots}
\end{figure*}

We are interested in determining the robustness of our classifier --- how well does our classifier perform with different types of data? 
We split the test set into batches, splitting the data by a given feature's distribution into four quantiles, ensuring that each quantile contains an equal number of objects. Then, we evaluate the accuracy for each quantile separately, only using the final set of features from the SFS (\autoref{sec:feature_selection}). Since this accuracy is now a measure of robustness, we call it the robustness score. This method is applied to all features from our ``super-set'' (as derived in \autoref{fig:dcpy_topfeats}). For the multi-class classification task, robustness scores for the Clark and Canberra metric have been visualized as a heatmap in \autoref{fig:robustness_plots}. We note that, although all features in the ``super-set'' of the 15 best features are not used by the classifier itself, we still use them to split our quantiles, before dropping the unused feature based on the distance metric. This helps us compare the robustness for different metrics.

\begin{figure*}
    \centering
    \includegraphics[width=0.75\textwidth]{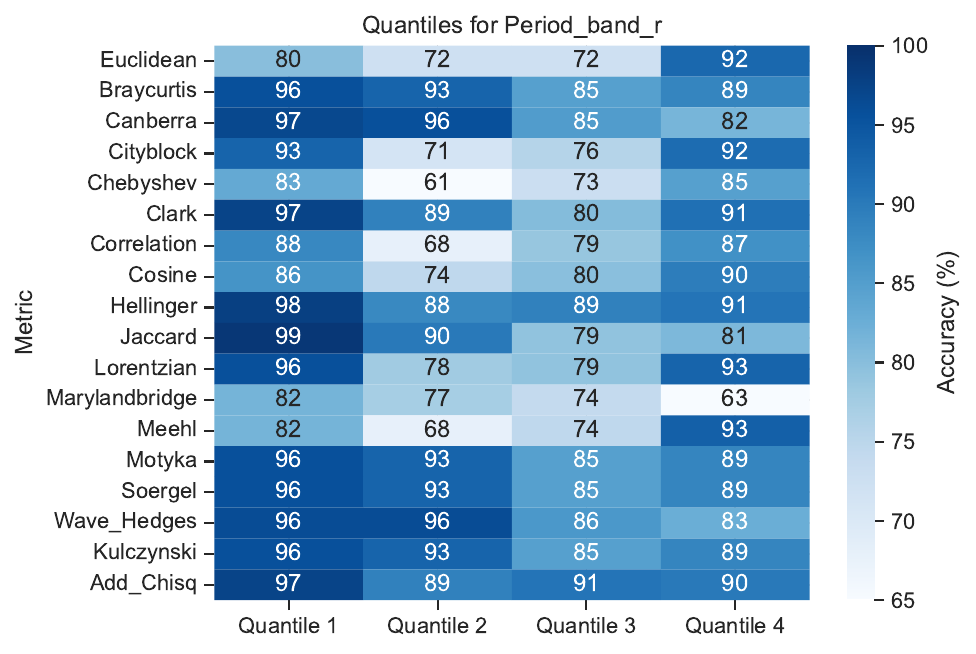}
    \caption{Robustness for objects with different periodicity (\texttt{Period\_band\_r}) for the multi-class classification task with the \dmc. The $r$-band period is the most important feature for most metrics. We compare how robust each of our metrics is to this feature. This figure is designed like \autoref{fig:robustness_plots}. For variables having a shorter period (and belonging to Quantile 1), the Jaccard metric gives the most accurate classification. On the other hand, for variables with a longer period (and belonging to Quantile 4), the Meehl metric gives the most accurate classification.}
    \label{fig:robustness_bestfeat}
\end{figure*}

In addition to obtaining the robustness scores for a given distance metric across all features, we can also compare the robustness of all distance metrics across a given feature. In \autoref{fig:robustness_bestfeat}, we look at the robustness score for the multi-class classification task across all the 18 distance metrics when splitting our test data based on \texttt{Period\_band\_r} (the $r$-band period, which is one of the most important features for the multi-class classification, as we saw earlier in \autoref{fig:dcpy_topfeats}).

The performance for different quantiles of the $r$-band period varies for different distance metrics. For example, the Jaccard metric provides a highly accurate prediction (99\%) for objects having short $r$-band periods (such that they belong in Quantile 1), but not so much (81\%; Q4) for objects with a long period (from Quantile 4). Conversely, the Meehl metric has a low accuracy (82\%; Q1) for objects with shorter periods, but has some of the best performance (93\%; Q4) for objects with longer periods. The fact that performance for different quantiles varies for different metrics may have remarkable implications. This can enable a dynamic approach for classification - when a test object is discovered, a different distance metric might be suitable based on what values its features have. 
This dynamism is something that cannot be achieved with traditional classifiers.

\subsection{Interpretability}
One more key advantage that \distclassipy\ offers is better interpretability compared to methods like RFC and Neural Networks. When a test object is classified using the \dmc, we know exactly what features are being used (unlike some deep learning methods), and we also know which features are more important. Furthermore, the decision making process is also fairly straightforward, unlike an RFC, whose constituent trees often have arbitrary decision boundaries. The \dmc\ essentially finds the class to which the test object is ``closest'', in a scaled feature-space. This can also be directly visualized for lower dimensions for a given metric space, and this may play a significant role in understanding a particular classification.

\section{Conclusions}
\label{sec:conclusions}

Time series classification is a non-trivial problem in machine learning. In
ground-based astronomy, light curve classification is even more challenging due
to the sparsity and heteroscedasticity of the data. We developed a \dmc\,
\distclassipy\ that uses distance metrics computed on features derived from
light curves. Tested on multiple classification tasks, our classifier produces
results comparable to other state-of-the-art feature-based methods like Random
Forest Classifiers (RFCs), with $F_1=92\%$ and $F_1=94\%$ on a multi-class
and a \ovr\ classification respectively. We achieve a performance of
$F_1=68\%$ on a particularly challenging binary classification that attempts
to separate RSCVN and BYDra, two star types that exhibit photometrically
similar and highly variable behaviors~\citep{bopp_1980}, and for which, too,
the RFC performance is comparable $F_1=70\%$. This cross-validated
performance was confirmed by running the model on a ``hidden'' dataset
completely unseen in training and that, unlike the original dataset, is not
rebalanced to have an even representation of all classes. For all three
classification tasks, we confirm the performance obtained in cross-validation
testing: $F_1 = 93\%, 92\%$, and 68\% respectively for the multi-class, \ovr, and
binary classification. 

At the same time, \distclassipy\ is faster, more interpretable, and suited to be tailored to specific science goals with additional performance enhancements and computational advantages. In our tests, it performed $50\times$ faster than RFC on small datasets (10 features by 100 objects), and over $150\times$ faster on larger datasets (500 features by 10,000 objects) demonstrating impressive scalability properties. The dynamism in the distance metrics allows for a choice suited to the available features and the test particle's characteristics within the feature space. 
Finally, because the features are a result of domain-sensitive dimensionality reduction, the user can leverage relevant knowledge in the context of the problem.

\distclassipy\ is made available to the users as open-source code. The package is accessible through PyPI\footnote{\url{https://pypi.org/project/distclassipy/}} with the goal of broadening its potential applications not only in astronomy but also in other classification scenarios across various fields.

In future work, we plan to extend \distclassipy\ to transient classification, explore its use for anomaly detection, and test more distance metrics, including metrics typically reserved for comparison of statistical distributions, \eg, Earth Mover's Distance, Bhattacharya distance. 

\section*{Acknowledgment}
SC would like to acknowledge the support received from Prof. Sukanta Panda and the Department of Physics, IISER Bhopal. This work originated as part of SC's Master's thesis at IISER Bhopal.

Based on observations obtained with the Samuel Oschin Telescope 48-inch and the 60-inch Telescope at the Palomar Observatory as part of the Zwicky Transient Facility project. ZTF is supported by the National Science Foundation under Grants No. AST-1440341 and AST-2034437 and a collaboration including current partners Caltech, IPAC, the Weizmann Institute for Science, the Oskar Klein Center at Stockholm University, the University of Maryland, Deutsches Elektronen-Synchrotron and Humboldt University, the TANGO Consortium of Taiwan, the University of Wisconsin at Milwaukee, Trinity College Dublin, Lawrence Livermore National Laboratories, IN2P3, University of Warwick, Ruhr University Bochum, Northwestern University and former partners the University of Washington, Los Alamos National Laboratories, and Lawrence Berkeley National Laboratories. Operations are conducted by COO, IPAC, and UW.

The authors acknowledge the support of the Vera C. Rubin Legacy Survey of Space and Time Science Collaborations\footnote{\url{https://www.lsstcorporation.org/science-collaborations}} and particularly of the Transient and Variable Star Science Collaboration\footnote{\url{https://lsst-tvssc.github.io/}} (TVS SC) that provided opportunities for collaboration and exchange of ideas and knowledge. 
The authors wish to thank the anonymous referee for insightful comments that lead us to improve our manuscript.
This research has made use of NASA’s Astrophysics Data System Bibliographic Services.

\section*{Data Availability}
Our dataset, along with all our code is available on \url{https://github.com/sidchaini/LightCurveDistanceClassification}. 

In addition, our classifier package \distclassipy\ is available on \url{https://pypi.org/project/distclassipy} and \url{https://github.com/sidchaini/DistClassiPy}

\section*{CRediT authorship contribution statement}
\textbf{Siddharth Chaini}: Conceptualization, Methodology, Software, Validation, Formal analysis, Investigation, Writing - Original Draft, Visualization. \textbf{Ashish Mahabal}: Conceptualization, Methodology, Formal analysis, Investigation, Resources, Writing - Review \& Editing, Supervision, Visualization. \textbf{Ajit Kembhavi}: Resources, Writing - Review \& Editing, Supervision. \textbf{Federica B. Bianco}: Conceptualization, Methodology, Investigation, Resources, Writing - Review \& Editing, Visualization, Supervision, Funding acquisition.

\section*{Declaration of Generative AI and AI-assisted technologies in the writing process}
During the preparation of this work the author(s) used ChatGPT to generate LaTeX code templates. After using this tool/service, the author(s) reviewed and edited the content as needed and take(s) full responsibility for the content of the publication.

\bibliographystyle{elsarticle-harv}
\bibliography{refs}

\clearpage
\appendix \label{appendix}

\section{Distance: A Mathematical Perspective} \label{app:metrics}

Earlier, in the paper, we defined the distance as:

\defdistance*

From the three axioms in \autoref{def:distance}, another condition (\emph{non-negativity}) can be shown to hold:

\begin{cor}
\begin{equation}
d(x,y) \geq 0, \text{for all } x,y \in X
\end{equation}
\end{cor}

\begin{exmp}
The Euclidean distance between two points in $\mathbb{R}^2$ is an example of a distance. Similarly, we can define a distance between any set of points such that \autoref{def:distance} is satisfied.
\end{exmp}

\begin{defn} \label{def:metric_space}
A metric space $(X,d)$ is a set $X$ equipped with a metric $d$.
\end{defn}

\begin{exmp}
The 2-dimensional $\mathbb{R}^2$ plane with the Euclidean distance is an example of a metric space.  Similarly, we can also define metrics on matrices, functions, sets of points or any other mathematical object as long as it satisfies~\ref{def:metric_space}.
\end{exmp}

Now, we note the difference between the terms \emph{distance} and \emph{distance metric} (often shortened to just \emph{metric}).
\begin{defn}
The term \emph{metric} refers to the way to calculate the distance between any two elements of the set. Meanwhile, the term \emph{distance} refers to the scalar value obtained after using a particular metric on a pair of elements in the set.
\end{defn}

\begin{exmp}
The Euclidean metric for $\mathbb{R}^2$ is given by $ d = \sqrt{(x_2 - x_1)^2 + (y_2 - y_1)^2} $, while the Euclidean distance between the points $(0,0)$ and $(1,1)$ is given by $ d = \sqrt{2} $.
\end{exmp}

\section{List of Metrics} \label{app:list_of_metrics}
We used the following metrics for all distance calculations in this paper. For
a detailed description of each metric,
see~\citet{dezaEncyclopediaDistances2013}.
\begin{enumerate}
    \item \textbf{Euclidean metric}:\\
    \begin{equation} \label{eq:euclidean_metric}
        d(u,v)=\sqrt{\sum \big(v_i - u_i\big)^2}
    \end{equation}
    The Euclidean metric is the most widely used metric and represents the length of the shortest straight line between two points in a multidimensional space.

    \item \textbf{Cityblock metric}:\\
    \begin{equation} \label{eq:cityblock_metric}
        d(u,v)={\sum\big|v_i - u_i\big|}
    \end{equation}
    The Cityblock metric gives the shortest path between 2 points if we restrict movement to the grid. Named after the grid layout of cities, this is the shortest distance for a taxi to traverse between two places. It is also known as the Manhattan metric.

    \item \textbf{Bray-Curtis metric}:\\
    \begin{equation} \label{eq:bray_curtis_metric}
        d(u,v)=\dfrac{\sum\left|u_{i}-v_{i}\right|}{\sum\left|u_{i}+v_{i}\right|}
    \end{equation}
    The Bray-Curtis metric is commonly used in Ecology and Biology, and denotes the dissimilarity in the (biological) species composition of two sites.

    \item \textbf{Canberra metric}:\\
    \begin{equation} \label{eq:canberra_metric}
        d(u, v)=\sum_{i} \frac{\left|u_{i}-v_{i}\right|}{\left|u_{i}\right|+\left|v_{i}\right|}
    \end{equation}
    The Canberra metric is a weighted version of the Cityblock metric, with a denominator that is slightly different from the Bray-Curtis metric. The form of the denominator makes this metric sensitive to points that are closer to the origin.

    \item \textbf{Chebyshev metric}:\\
    \begin{equation} \label{eq:chebyshev_metric}
        d(u,v)=\max\big(|u_i-v_i|\big)
    \end{equation}
    The Chebyshev metric is also known as the chessboard metric. It represents the minimum number of moves needed by the King chess piece to move from one point to another.

    \item \textbf{Clark metric}:\\
    \begin{equation} \label{eq:clark_metric}
        d(u,v)=\left(\frac{1}{n} \sum\left(\frac{u_{i}-v_{i}}{\left|u_{i}\right|+\left|v_{i}\right|}\right)^{2}\right)^{\frac{1}{2}}
    \end{equation}
    The Clark metric is a variation of the Euclidean metric, but with a normalizing factor similar to the Canberra metric. 
    
    \item \textbf{Cosine metric}:\\
    \begin{equation} \label{eq:cosine_metric}
    \begin{aligned}
        & d(u,v) = \\
        &\frac{\sum\left((u_{i}-\bar{u})\cdot(v_{i}-\bar{v})\right)}{\sqrt{\sum\left((u_{i}-\bar{u})\cdot(u_{i}-\bar{u})\right)}\sqrt{\sum\left((v_{i}-\bar{v})\cdot(v_{i}-\bar{v})\right)}}
    \end{aligned}
    \end{equation}
    The Cosine metric is based on the cosine similarity, which depends only on the angle between the two vectors in the feature space, and not their magnitude.

    \item \textbf{Hellinger metric}:\\
    \begin{equation} \label{eq:hellinger_metric}
        d(u,v)=\sqrt{2 \sum\left(\sqrt{\frac{u_{i}}{\bar{u}}}-\sqrt{\frac{v_{i}}{\bar{v}}}\right)^{2}}
    \end{equation}
    The Hellinger metric is a metric that is commonly used to compare two probability distributions but can also be used with numeric data.

    \item \textbf{Jaccard metric}:\\
    \begin{equation} \label{eq:jaccard_metric}
        d(u,v)=\frac{\left(\sum u_{i}-v_{i}\right)^{2}}{\sum u_{i}^{2}+\sum v_{i}^{2}-\sum u_{i} v_{i}}
    \end{equation}
    The Jaccard metric is based on the Jaccard similarity coefficient, which is a measure of similarity between 2 sets. The similarity coefficient can be converted into a distance.

    \item \textbf{Lorentzian metric}:\\
    \begin{equation} \label{eq:lorentzian_metric}
        d(u,v)=\sum \ln \left(1+\left|u_{i}-v_{i}\right|\right)
    \end{equation}
    Because the Lorentzian metric is given by a natural log, it is more senstive to changes when the input vector is small, but less sensitive to changes when the input vector is large.

    \item \textbf{Meehl metric}:\\
    \begin{equation} \label{eq:meehl_metric}
        d(u,v)=\sum_{1 \leq i \leq n-1}\left(u_{i}-v_{i}-u_{i+1}+v_{i+1}\right)^{2}
    \end{equation}
    The Meehl metric mixes terms from consecutive dimensions, thus making the ordering of the dimensions significant.

    \item \textbf{Soergel metric}:\\
    \begin{equation} \label{eq:soergel_metric}
        d(u,v)=\frac{\sum\left|u_{i}-v_{i}\right|}{\sum \max \left\{u_{i}, v_{i}\right\}}
    \end{equation}
    The Soergel metric is another modified version of the Cityblock metric but has only a $\max$ term in the denominator. Note that the summation is performed independently on the numerator and denominator before division.

    \item \textbf{Wave–Hedges metric}:\\
    \begin{equation} \label{eq:wave_hedges_metric}
        d(u,v)=\sum \frac{\left|u_{i}-v_{i}\right|}{\max \left\{u_{i}, v_{i}\right\}}
    \end{equation}
    The Wave-Hedges metric is a slight modification of the Soergel metric, where the summation is done after division.

    \item \textbf{Kulczynski metric}:\\
    \begin{equation} \label{eq:kulczynski_metric}
        d(u,v)=\frac{\sum\left|u_{i}-v_{i}\right|}{\sum \min \left\{u_{i}, v_{i}\right\}}
    \end{equation}
    The Kulczynski metric is similar to the Soergel metric, with the difference that its denominator consists of a minimum instead of a maximum.

    \item \textbf{Additive Symmetric $\chi^2$ metric}:\\
    \begin{equation} \label{eq:add_chisq_metric}
        d(u,v)=\Sigma \dfrac{(u_i-v_i)^2 (u_i+v_i)}{u_i v_i}
    \end{equation}
    The additive symmetric $\chi^2$ metric is a form of a weighted Euclidean metric, where terms between the two vectors are mixed.

    \item \textbf{Correlation metric}:\\
    \begin{equation} \label{eq:correlation_metric}
        d(u,v)=1-\frac{(u-\bar{u}) \cdot(v-\bar{v})}{\|(u-\bar{u})\|_{2}\|(v-\bar{v})\|_{2}}
    \end{equation}
    The Correlation metric is based on the Pearson correlation coefficient between two vectors.

    \item \textbf{Maryland Bridge metric}:\\
    \begin{equation} \label{eq:maryland_bridge_metric}
        d(u,v)=1-\frac{1}{2}\left(\frac{\sum u_{i} v_{i}}{\sum u_{i}^{2}}+\frac{\sum u_{i} v_{i}}{\sum v_{i}^{2}}\right)
    \end{equation}
    The Maryland Bridge metric is based on the Maryland Bridge similarity.
    This metric does not satisfy axiom 3 (triangle inequality) in
    \autoref{def:distance} and it is therefore not a distance metric. However, it has
    been used in literature, particularly in the field of  genomics to measure
    similarity between genomes~\citep{mirkin2003top}. Thus, it is included as a
    utility in our package.
    \item \textbf{Motyka metric}:\\
    \begin{equation} \label{eq:motyka_metric}
        d(u,v)=\frac{\sum \max \left\{u_{i}, v_{i}\right\}}{\sum\left(u_{i}+v_{i}\right)}
    \end{equation}
    The Motyka metric has a summation term in the denominator, which acts as a scaling factor of the maximum term in the numerator.
    This metric does not satisfy axiom 1 (identity of indiscernibles) in \autoref{def:distance} (and it is not positive defined). However, it is a metric that commonly appears in the literature when measuring similarity between data, so it is included as a utility in our package.
\end{enumerate}

\section{Hardware and Software specifications} \label{app:specs}

The results described in this work, including the computational time estimates, were produced on machines with the hardware specifications reported in \autoref{tab:hardware}. The Python libraries and corresponding versions used in this work are given in \autoref{tab:software}.

\begin{table}[h]
    \centering
    \resizebox{\columnwidth}{!}{
    \begin{tabular}{ll}
        \hline
    \textbf{Specification} & \textbf{Details} \\
        \hline
        Processor & ARM \\
        CPU & Apple M1 Pro \\
        Physical cores & 8 \\
        Total cores & 8 \\
        RAM & 16 GB \\
        Operating System & Darwin 23.1.0 \\
        Machine & arm64 \\
        Platform & macOS-14.1.1-arm64-arm-64bit \\
        \hline
    \end{tabular}
    }
    \caption{Hardware Information}
    \label{tab:hardware}
\end{table}

\begin{table}[h]
    \centering
    \resizebox{\columnwidth}{!}{
    \begin{tabular}{lll}
        \hline
    \textbf{Library} & \textbf{Version} & \textbf{Reference}\\
        \hline
        Python & 3.12.2 & ~\citet{python3}\\
        \texttt{\distclassipy} & 0.1.5 & ~\citet{2024ascl.soft03002C}\\
        \texttt{NumPy} & 1.26.4 & ~\citet{numpy}\\
        \texttt{Pandas} & 2.2.1 & ~\citet{pandas,pandas2}\\
        \texttt{scikit-learn} & 1.4.1.post1 & ~\citet{scikit-learn}\\
        \texttt{SciPy} & 1.12.0 & ~\citet{scipy}\\
        \texttt{statistical-distances} & 0.9.1 & ~\citet{zielezinskiStatisticaldistanceDistanceMeasures}\\
        \texttt{matplotlib} & 3.8.3 & ~\citet{matplotlib}\\
        \texttt{JSON} & 2.0.9 & ~\citet{json}\\
        \texttt{tqdm} & 4.65.0 & ~\citet{tqdm}\\
        \texttt{seaborn} & 0.13.2 & ~\citet{seaborn}\\
        \texttt{mlxtend} & 0.23.1 & ~\citet{raschkaMLxtendProvidingMachine2018}\\
        \texttt{Jupyter} & 4.1.2 & ~\citet{jupyter}\\
        \hline
    \end{tabular}
    }
    \caption{Software and Versions}
    \label{tab:software}
\end{table}

\end{document}